# Cell Balancing for the Transportation Sector: Techniques, Challenges, and Future Research Directions

Anupama R Itagi[1], Rakhee Kallimani[2], Krishna Pai[3], Sridhar Iyer[4], and Onel L. A. López[5]

[1]Department of Electrical and Electronics Engineering,
KLE Technological University, 580032, Hubballi, Karnataka, India
Email: anupama_itagi@kletech.ac.in

[2]Department of Electrical and Electronics Engineering,
KLE Technological University Dr. MSSCET, 590008, Belagavi, Karnataka, India
Email: rakhee.kallimani@klescet.ac.in

[3]Independent Researcher,
560094, Bengaluru, Karnataka, India
Email: krishnapai271999@gmail.com

[4]Department of CSE(AI),
KLE Technological University Dr. MSSCET, 590008, Belagavi, Karnataka, India
Email: sridhariyer1983@klescet.ac.in

[5]Faculty of Information Technology and Electrical Engineering,
University of Oulu, 90014, Oulu, Finland
Email: onel.alcarazlopez@oulu.fi

**Abstract**

Efficient and reliable energy systems are key to progress of society. High performance batteries are essential for widely used technologies like Electric Vehicles (EVs) and portable electronics. Additionally, an effective Battery Management System (BMS) is crucial to oversee vital parameters of battery. However, BMS can experience cell imbalance due to charging/discharging dynamics, which reduce battery capacity, lifespan, and efficiency, and raise critical safety concerns. This calls for effective cell-balancing techniques. Notably, the existing literature on cell balancing is limited, urgently necessitating a thorough survey to pinpoint key research gaps and suggest prompt solutions. In this article, cell balancing and corresponding techniques are reviewed. Initially, we detail comparison of passive cell balancing techniques and assess their respective advantages, drawbacks, and practical applications. Then, we discuss the strengths and weaknesses of active cell balancing methods and applicability of cell balancing for both, series- and parallel-connected cells. Additionally, we examine the need for cell balancing in commonly used batteries, and applications in EVs. Lastly, we present detailed prospects which include challenges and directions for future research.

## I. INTRODUCTION

Achieving sustainable urban development necessitates integrated planning, smart infrastructure investments, green space preservation, inclusive development, climate resilience, technological innovation, promotion of sustainable lifestyles, and long-term commitment. These aspects are reinforced by widespread adoption of Renewable Energy Sources (RES), which is gaining significant importance owing to depletion of fossil fuels and increasing demands for energy [1]. Therefore, RES is considered a fundamental pillar for sustainable practices [2]. Further, adopting Electric Vehicles (EVs) fuelled by clean energy derived from RES signifies a holistic approach towards achieving a more sustainable and environmentally conscious future [3]. This integration will address the collective challenge of reducing carbon footprints and advancing technology to meet the evolving needs of the planet which continuously searches for responsible energy solutions [4].

The EVs rely on batteries to power their electric drive trains. These batteries influence significantly the range and overall efficiency of the EVs, making them a key component in the shift towards sustainable and eco-friendly transportation [5]. The widespread use of batteries across sectors underscores the importance of detailed Battery Management System (BMS) analyses to enhance lifetime and performance of battery. The BMS is essential to any battery-operated supply arrangement [6], and serves as an essential guardian of a battery pack, responsible for maintaining its performance and safety. In general, as shown in Fig. 1, BMS comprises monitoring, controlling, and protection units. Here, equalisation is a crucial part of controlling units as it has a direct impact on the battery framework's durability. For operation, devices to monitor and control manage battery functions within safety margins bounds and offer protection from conditions which adversely affect the operation. BMS also monitors essential parameters such as current, voltage, temperature, State of Health (SoH), State of Charge (SoC), and State of Function (SoF) [7]–[10]. By keeping a close eye on these parameters, BMS ensures that operation of battery is within limits of safety and performance is optimised [7].

It is known that a battery full pack's ability reduces rapidly with continuous operations, which degrades the complete battery framework. This condition worsens with series connection of additional cells within battery, necessitating frequent charging through the battery string. In effect, this results in cell imbalance during the charge/discharge, which is key problem in BMS. Cell imbalance refers to the uneven distribution of charge or energy among individual cells within pack of battery [11]. Cell imbalance can arise due to intrinsic or extrinsic factors. Intrinsic factors are linked to manufacturing processes, e.g., the self-discharge rate can cause variations in active material and internal resistance [6], [12]. On the other hand, extrinsic factors encompass aspects such as connection configuration, charge and discharge current, and temperature [13]. The key factors responsible for cell imbalance [14] are shown in Fig. 2. Also, existing studies have shown that neglecting the key aspect of cell imbalance leads to adverse effects such as [15], [16]:

- Reduced Capacity: In the case of a series connection, cells with higher voltages tend to reach their maximum capacity earlier than others, limiting battery pack's capacity. This results in less energy storage and shorter runtime in applications such as EVs and portable electronics [17], [18]
- Reduced Efficiency: Imbalanced cells result in decrease of overall system efficiency as battery pack is prematurely discharged or charged due to the presence of overcharged or undercharged cells.
- Reduced Lifespan: Cells that are consistently operated at extreme voltage levels, either too high or too low, can experience early degradation, leading to a reduced battery lifespan.
- Safety Concerns: Extreme voltage differences among the cells can pose safety hazards. Specifically, overcharged cells may become unstable and could be prone to thermal runaway or cell rupture, potentially leading to fires or explosions.

Cell balancing has emerged as a key technique to improve battery safety and extend battery life [19]. Currently, researchers are investigating methods to enhance the performance of cell balancing circuit within BMS control unit. However, research on cell balancing is in the early stages and key factors contributing to cell imbalance are still being identified. Further, although initial research has demonstrated the advantages of implementing cell balancing techniques in BMSs, there is still a limited understanding of how cell balancing contributes to the overall system performance. The major questions which warrant solutions are the: (i) applicability of cell balancing to both, series- and parallel-connected cells, (ii) merits and demerits of implementing active and passive cell balancing, and (iii) influence of battery chemical properties on the effectiveness of cell balancing techniques.

In view of the above research gaps, this article provides an in-depth overview, listing numerous unresolved research questions and challenges, along with potential directions for effective solutions.

TABLE I
THE LIST OF IMPORTANT ACRONYMS

| Acronym | Definition | Acronym | Definition |
| --- | --- | --- | --- |
| BMS | Battery Management System | EVs | Electric Vehicles |
| RES | Renewable Energy Source | SoH | State of Health |
| SoC | State of Charge | SoF | State of Function |
| SoE | State of Energy | BTMS | Battery Thermal Management Systems |
| Pb-acid | Lead acid | Ni-Cd | Nickle-Cadmium |
| Ni-MH | Nickel–metal hydride | Li-ion | Lithium-ion |
| Li-polymer | Lithium-polymer | PWM | Pulse-Width Modulation |
| MLCs | Multi-Level Converters | ML | Machine Learning |
| DL | Deep Learning | LIB | Lithium-ion Battery |
| MOSFET | Metal-Oxide-Semiconductor Field-Effect Transistor | BESS | Battery Energy Storage System |
| ADC | Analog to Digital converter | BPNN | Back Propagation Neural Network |
| OCV | Open Circuit Voltage | MPC | Model Predictive Control |
| DPS | Dual Phase-Shift | ICB | Integrated Cascaded Bidirectional |
| EESS | Electrical Energy Storage System | AC2AC | Any Cell to Any Cell |
| DC2C | Direct Cell to Cell | EoL | End of Life |
| ICB | Integrated Cascaded Bidirectional | DAB | Dual active bridge converter |
| CIDAB | Capacitively Isolated Dual Active Bridge | SPST | Single-Pole Single-Throw |
| DPDT | Double Pole Double Throw | PHEV | Plug-in Hybrid Electric Vehicles |
| AER | All-Electric Range | HEVs | Hybrid Electric Vehicles |
| ISO | International Standardisation Organisation | JIS | Japanese Industrial Standard |
| IEC | International Electrochemical Commission | SAE | Society of Automotive Engineers |
| PV | Photovoltaic | AI | Artificial Intelligence |
| IoT | Internet of Things | SDG | Sustainable Development Goals |

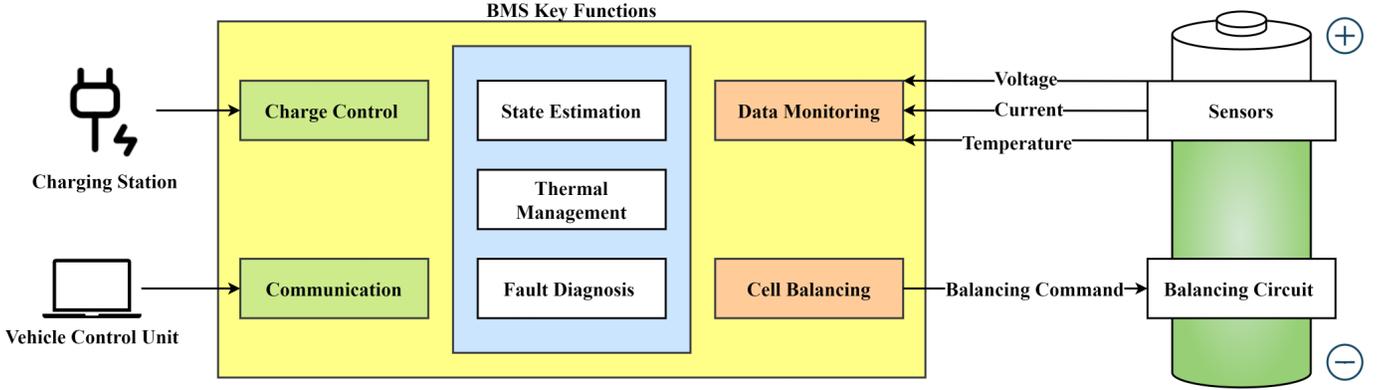

Fig. 1. Key functions of BMS [13]

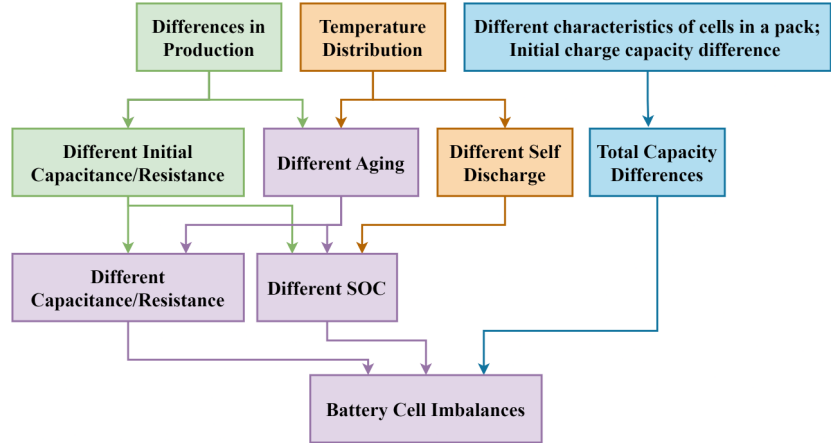

Fig. 2. The key factors affecting cell imbalance [14]

*A. Research Methodology*

Herein, we overview the literature related to cell balancing which sets the survey's context and aids in identifying major research gaps. Specifically, the survey is conducted following guidelines laid out in [20]–[23] which enables framing and responding to following research questions:

- Do the techniques contributing to cell balancing enhance BMS performance and to what extent?
- What are the merits and demerits of implementing active/passive methods of cell balancing? Specifically, will the implementation depend on number of elements and equalisation configuration?
- Is cell balancing applicable to both series- and parallel-connected cells? A follow-up question is regarding the considerations while implementing cell balancing in each configuration

From our detailed survey, we identify that cell balancing is necessary for commonly used types of battery. However, impact of cell balancing is different for each type when used in BMS.

The survey is conducted in three stages to (i) identify keywords and define strings for searching, (ii) choose sources for data, and (iii) search within sources of data. The search is conducted over online sources viz., Google Scholar, SCOPUS, WoS, and specific journal sites (i.e., IEEE, Elsevier, MDPI, Wiley-Blackwell, and Springer) as databases. Additionally, search of manual type is also conducted over authors' profiles which fall withing the domain of cell balancing. The keywords for search include BMS, battery monitoring, cell balancing, SoC estimation, battery health monitoring, and battery fault diagnosis. The manner in which an individual database contributed to the survey is shown in Fig. 3 (a). The search at the start resulted in 230 aggregate sources from which, a through review and removal of duplicates resulted in 173 sources. Thereafter, following the criteria which includes and excludes resulted in a reduction of further 41 sources. To include a source, the followed criteria targets (i) research focus of the source on BMS and cell balancing, and (ii) publication of the source in major database between years 1999-2024. Overall, an aggregate of 132 sources are included amongst which, as shown in Fig. 3 (b), 81 are primary studies, 39 are proceedings and/or reviews, and 12 are technical reports.

Next, content analysis is implemented to extract data and conduct synthesis [24]. This process aids in synthesising data as per cell balancing elements detailed in Section II, Section IV, and Section V. Finally, to access quality, each source is weighted over following questions [21]: (i) are goals of research clear, (ii) are the aims attained, and (iii) are the process for research and related findings appropriate. This could result in a source answering (i) all questions - 1; (ii) all questions partially - 0.5; and (iii) no questions - 0. While performing this quality classification for the current article, it was observed that all the primary sources responded with a '1', and hence, there were no exclusions. Further, included sources enabled in extracting the (i) major use-case of cell balancing detailed in Section IV), and (ii) challenges detailed in Section V).

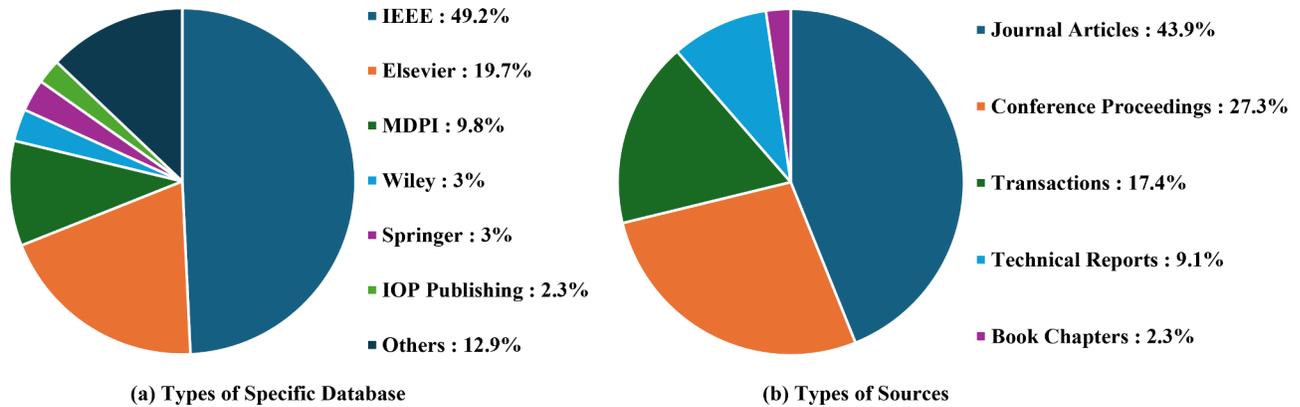

Fig. 3. (a) Studies included from specific databases. (b) Types of included sources.

*B. Recent Surveys on Cell Balancing*

The methodology detailed in previous sub-section helped to pinpoint recent articles that review and analyze different facets of BMS and cell balancing. The reviewed articles cover topics such as cell balancing techniques for various types of batteries, Battery Thermal Management Systems (BTMS), utilisation of intelligent cloud computing techniques in BMS, and fault diagnostics and prognostics. Next, we detail the various aspects of these recent articles.

A comprehensive review of BMS is conducted in [25] with a specific focus on evaluating different cell-balancing technologies that are specifically tailored for integration into EVs. The article delves into the intricate workings of BMS, shedding light on its various components, functionalities, and operational mechanisms. Through meticulous analysis, the paper identifies and assesses a range of cell-balancing techniques that exhibit compatibility and efficacy in the context of EV applications. The authors in [26] provide an in-depth analysis of BTMS focusing on its design optimisation technologies and multi-objective optimisation approaches. Through extensive exploration, the article delves into the capabilities of BTMS in minimising the peak battery temperatures and spatial temperature differentials, which are crucial factors affecting battery performance and longevity. By discussing various optimisation techniques and strategies, the article highlights the effectiveness of BTMS in maintaining optimal battery operating conditions, thereby enhancing overall system efficiency and reliability. Through this comprehensive examination, the article offers valuable insights into the intricate workings of BTMS and its role in mitigating thermal challenges associated with battery systems, ultimately contributing to advancements in energy storage technology.

A thorough examination of battery balancing configurations, including their control strategies, security measures, and practical applications, is provided in [27]. However, the article does not highlight any case studies showcasing the implementation of cell balancing techniques and possible solutions for future research challenges. Advancements in power management systems and strategies for battery charging control are discussed in [28]. The article emphasises on leveraging intelligence-based cloud computing technologies to improve battery longevity and efficiently regulate SoC. This integration enables the monitoring and optimisation of battery performance, ensuring optimal operation and maximising overall efficiency. The authors in [29] offer an exhaustive examination of contemporary cell balancing techniques specifically designed for low voltage applications. The study considers crucial factors including control complexity, switch stress, cell balancing speed, cost implications, and circuit size.

The complexities of SoC balancing within battery storage systems utilising Multi-Level Converters (MLCs) are discussed in [30]. The article encompasses a broad spectrum of issues spanning from conventional technologies to recent advancements in the field. The authors in [31] present a comprehensive overview of techniques implemented for SoC estimation in battery packs, examining their efficacy and the impact of cell inconsistencies on pack performance and SoC estimation accuracy. Additionally, the article discusses a range of cell balancing techniques necessary to mitigate these inconsistencies. By exploring the relationship between SoC estimation techniques and cell balancing strategies, the study contributes to a deeper understanding of the interplay between these two critical aspects of battery management. The study however does not present details of standardised dataset, and SoC evaluation techniques. The study in [32] provides valuable insights into the common battery faults, explores the potential of utilising deep learning (DL) techniques for diagnosing and prognosticating EV battery faults, and discusses state-of-art DL technologies implemented for fault detection. The study offers critical perspectives on improving the reliability and performance of the EV battery systems.

The authors in [7] conducted a thorough examination of fault modes, fault data, and fault diagnosis techniques across various scenarios, encompassing laboratory settings, EVs, energy storage systems, and simulations. Through extensive analysis, the authors compared the suitability and effectiveness of different techniques across diverse contexts, shedding light on their performance and limitations. Furthermore, the study explores characteristics of fault data, technique performance, and inherent limitations, providing a comprehensive understanding of fault diagnosis in different applications. The study in [33] offers a comprehensive review of models in Lithium-Ion Battery (LIB) risk management focusing on both, electrical and thermal behaviours of the batteries. The article discusses systematic techniques applied to LIB-related accidents, providing insights into the root cause and progression of thermal runaway. Through detailed analysis, the article provides valuable insights into risk management strategies for LIBs, facilitating the development of effective preventive measures and enhancing the safety and reliability of battery systems.

It is worth highlighting that all the above recent articles do not include (i) any case studies to showcase the implementation of cell balancing techniques and (ii) potential avenues for future research. In comparison, in the current survey, we present a transportation case study, and detail possible future avenues for research. Table II compares this article's contribution with the recent articles on cell balancing.

TABLE II
SUMMARY OF MOST RECENT SURVEYS RELATED TO CELL BALANCING

| References | Main Contributions | Constraints related to our survey |
|---|---|---|
| [7] | • Discuss fault modes, related data, and relevant diagnosis techniques across different scenarios, including laboratory settings, electric vehicles, energy storage systems, and simulations.<br>• Extensive analysis of techniques, comparing their suitability and effectiveness across diverse contexts is discussed.<br>• Explore fault data characteristics, technique performance, and inherent limitations for a comprehensive understanding. | Future research directions are not articulated. |
| [25] | Explore details of BMS and evaluate various cell-balancing technologies suitable for implementation in EVs. | • Lacking a case study demonstrating the implementation of cell balancing techniques.<br>• Future research directions are not articulated. |
| [26] | Extensively discuss Battery Thermal Management System (BTMS), the capabilities of design optimisation technologies, and multi-objective optimisation approaches to minimise peak battery temperatures and spatial temperature differentials. | Lack of future research challenges and possible solutions. |
| [27] | Extensively review battery balancing configurations, their control strategies, security, and applications. | • Lacking a case study on the implementation of cell balancing techniques.<br>• Future research directions are not articulated. |
| [28] | • Delve into contemporary literature exploring advanced systems for managing power distribution and optimizing charging control of battery technologies.<br>• Emphasise the integration of cutting-edge intelligent cloud computing technologies to elevate battery longevity and effectively control SoC. | Future research directions are not articulated. |
| [29] | • Offers an extensive examination of various cell balancing methodologies specifically customised for low-voltage applications.<br>• Factors such as control complexity, balancing speed, switch stress, circuit size and cost are taken into account. | • Lacking a case study demonstrating the implementation of cell balancing techniques.<br>• Future avenues for research are not articulated. |
| [30] | SOC balancing within Multi-Level Converter (MLC) based battery storage systems is discussed encompassing both, conventional approaches and recent advancements in the field. | • Lacking a case study demonstrating the implementation of cell balancing techniques.<br>• Future research directions are not articulated. |
| [31] | • Provide an overview of battery pack SoC estimation techniques, exploring their effectiveness and the influence of cell inconsistencies on pack functioning and SoC approximation accuracy.<br>• Various battery cell equalisation techniques required to address these inconsistencies are discussed. | • No standardised dataset and SoC evaluation technique.<br>• Future research directions are not articulated. |
| [32] | • Provides information on typical battery faults.<br>• The potential of utilising Deep Learning (DL) in diagnosing and predicting faults in EV batteries is discussed.<br>• Discusses contemporary DL methodologies applied to detect faults. | Demonstration of test cases not discussed |
| [33] | • Present a comprehensive review of models in LIB risk management focusing on the electrical and thermal behaviours of batteries.<br>• Systematic techniques applied to LIB-related accidents are discussed<br>• Provide insights into the origins and development of thermal runaway. | Future avenues for research are not articulated. |
| Our Survey | • Comparison of passive cell balancing techniques, and assessment of their respective advantages, drawbacks, and practical applications.<br>• Scrutinise the strengths and weaknesses of active cell equalisation techniques.<br>• Delve on the applicability of cell equalisation techniques for both, series and parallel connected cells.<br>• Examine whether cell balancing is essential for all battery types. | - |

*C. Current article's contribution*

From the previous sub-section it is clear that although there exists a extensive literature addressing cell balancing, the following aspects are not thoroughly addressed: (i) comparison and assessment of passive cell balancing techniques, (ii) scrutiny of strengths and weaknesses of active cell balancing techniques, and (iii) applicability of cell balancing for series and parallel connected cells. Compared to recent studies, survey in this article details the comparison of passive cell balancing techniques, and assessment of their respective advantages, drawbacks, and practical applications. Further, we also scrutinise the strengths and weaknesses of active cell balancing techniques, and delve into applicability of cell balancing for both, series and parallel-connected cells. Additionally, we examine whether cell balancing is essential for all battery types. Lastly, this survey discusses major research challenges and proposes possible directions.

The article's main contributions are as follows:
- We explore the benefits and identify the factors contributing to cell imbalancing in battery systems.
- We scrutinise the strengths and weaknesses of passive and active cell balancing techniques in detail.
- We delve into the applicability of cell balancing for both, series and parallel-connected cells.
- We examine the need for cell balancing in various battery types, and investigate the impact of battery chemical properties on cell balancing.

A list of the important acronyms used throughout the article is presented in Table. I, and the rest of the paper is structured as follows. Section II discusses cell balancing by considering battery technologies, the operation, and the techniques for cell balancing. In Section III, we overview the most recent studies that target multiple issues in BMS and cell balancing. Section IV discusses the transportation sector, which has emerged as the most popular application of cell balancing. Section V identifies major research challenges and proposes possible directions. Finally, Section VI concludes the survey.

## II. Fundamentals of Cell Balancing

Various types of batteries are documented in the literature. However, many are still in the research phase and not yet commercialised due, for instance, to immaturity, low energy density, safety concerns, high toxicity levels, and pricing [34].Batteries are commonly categorised as primary and secondary types. Primary batteries are designed for single-use purposes and incapable of recharging due to irreversible chemical reactions upon complete discharge, while secondary batteries are rechargeable, rendering them more suitable for applications in automotive and transportation sectors. Hence, rechargeable batteries such as Lead acid (Pb-acid), Nickle-Cadmium (Ni-Cd), Nickel–Metal Hydride (Ni-MH), Lithium-ion (Li-ion), and Lithium-polymer (Li-polymer), are mostly commercially used.

Among the available batteries, Pb-acid batteries stand out due to affordability. However, these batteries are plagued by short lifespans, high maintenance requirements, and environmental hazards due to toxic components. Ni-Cd batteries lack distinct advantages and are burdened by their weight, low energy density, and toxic composition. Ni-MH batteries offer superior specific energy but suffer from a shorter life cycle. Li-ion batteries provide specific energy, power, and rapid charging, but raise safety concerns. Li-polymer batteries are cost-effective and provide higher energy density with the additional benefit of leak-proof solid-state design. Flow batteries offer simplicity but face hurdles in transportation applications due to high electrolyte storage needs and volume-weight restrictions. The selection of a battery relies on particular requirements of the application(s), and consideration of key factors such as cost, energy density, charging speed, replacement frequency, operating temperature and voltage, and current requirements [1], [34], [35]. For example, Pb-acid batteries are mostly used in smart grid systems, whereas NiMH and Li-ion batteries are most suitable for transportation purpose [34]. However, due to their elevated specific energy and cost-effectiveness, Li-ion batteries are preferred over NiMH batteries for automotive applications [36]. Ni-Cd and Li-polymer batteries are commonly utilised in portable electronic devices that require lower current requirements [1]. Flow batteries find applications in medium and large-scale grid applications. In Table. III, we provide a comprehensive overview of the pros and cons associated with the various battery types which are commonly utilised across different applications.

Cell balancing refers to equalising the distribution of charge or energy among individual cells within a battery pack [29], [42]. Maintaining uniform voltage and SoC across battery cells is crucial during both the charging and discharging processes. The cell equalisation process helps to maximise the battery pack's overall functionality, lifespan, and protection against hazards [13], and its efficient implementation provides many advantages such as (i) provisioning solution to thermal run away in batteries, (ii) ensuring safe and efficient operation of the batteries, (iii) extending life span of the battery considerably, (iv) using decommissioned batteries as energy storage devices in micro-grids [43] resulting in economic and environmental benefits [44], and (v) extending the range of the EVs to foster user acceptance [45]. Further, cell balancing is essential for certain battery types and applications to ensure optimal performance and safety, and it may not be universally required for all battery types. The decision to implement cell balancing should be based on a thorough understanding of the following factors:

- **Chemistry**: In general, the need for cell balancing in battery systems is influenced by various factors inherent to the battery chemistry. One key factor to consider is the voltage behaviour of the cells within the battery. Different cell chemistry exhibits unique voltage profiles during charging and discharging cycles, and variations in these profiles can lead to cell imbalances [46]. Batteries such as Li-ion that experience large differences in voltage levels among cells due to factors such as manufacturing variances or aging effects, necessitate cell balancing to maintain uniform voltage across all the cells [47]. The internal impedance variances and discrepancies in the self-discharge rates also contribute significantly to the cell imbalance [36].

TABLE III
COMPARISON OF DIFFERENT BATTERY TECHNOLOGIES [1], [34], [35], [37]–[41]

| Battery type | Merits | Demerits | Applications |
|---|---|---|---|
| Pb-Acid | Low cost | • Typically lasts for three to five years.<br>• High regular maintenance, including replenishing with distilled water and cleaning terminals.<br>• Substantial weight.<br>• Limited energy density, constraining effectiveness in high-power applications.<br>• Environmental concerns arise from the presence of toxic materials such as lead, sulfuric acid, and cadmium.<br>• Restricted charging efficiency.<br>• Temperature sensitivity necessitating supplementary cooling or heating systems for optimal functionality. | • Energy storage device in smart grids.<br>• Emergency lighting |
| Ni-Cd | Long shelf life | • Low specific energy.<br>• Not suitable for applications that demand high instantaneous currents.<br>• Contains toxic metals such as lead, mercury, and cadmium.<br>• Lower cycle life | Portable electronic devices |
| Ni-MH | Higher specific energy | Lower cycle life. | Transportation application |
| Li-ion | • High specific energy.<br>• High specific power.<br>• Offers faster charging and discharging rates compared to Pb-acid and Ni-MH batteries.<br>• Longer life, provided proper conditions can be ensured.<br>• Higher Energy density. | Safety issues | • Transportation application. • Portable devices like mobile phones, laptops etc. |
| Flow batteries | • Provides enhanced operational flexibility by allowing easy control of cell-stack temperatures through external liquid heating or cooling mechanisms.<br>• Scalability<br>• Nearly unlimited longevity | • Not suitable for transportation due to high electrolyte storage needs.<br>• Enormous vessels required to accommodate high-capacity electrolyte.<br>• Volume and weight limitations for vehicle applications. | • Energy storage device in Medium to large scale grid.<br>• UPS |
| Li-polymer | • Lower cost<br>• Higher Energy density.<br>• Due to its solid-state nature, materials do not leak even in the event of an accident.<br>• Can be manufactured in any size or shape. | Performance is lower compared to Li-ion batteries. | Consumer electronics |

- **Configuration**: The configuration of batteries in series, parallel, or series-parallel connections can influence cell imbalance. Batteries connected in series are more prone to cell imbalance due to variations in the individual cell characteristics [36], whereas parallel connections may naturally balance the cells through shared current paths. Nevertheless, minor SoC fluctuations of the battery from circulating current require some attention [48].
- **Application**: The demands of the application in which the battery is utilised also hold significant importance. Cell balancing may be essential for critical applications where safety and reliability are paramount, such as EVs or grid energy storage systems, to prevent overcharging, undercharging, or thermal runaway [49]. In contrast, cell balancing may not be as critical for applications such as low-power consumer electronics.

The Cell equalisation can be executed through either voltage-based or SoC based methodologies [11], [50]. The voltage of a battery cell is impacted not only by the SoC but also notably influenced by the current. Consequently, a battery with an elevated voltage may possess a poorer SoC due to disparities in their current levels [51]. Since battery balancing primarily involves adjusting the cell's current, this results in varying voltages among batteries with the identical SoC during the equalisation process. Hence, using voltage as the indicator may result in sub-optimal equalisation outcomes [52]. Thus, SoC-based techniques are frequently preferred in contemporary approaches. The comparison of the aforementioned techniques is tabulated in Table. IV. Battery cell's SoC heavily relies on its electrochemical characteristics. The mutual dependence of key battery parameters is shown in Fig. 4. Variations in these characteristics across cells can lead to discrepancies in battery pack SoCs [30].

TABLE IV
COMPARISON OF VOLTAGE BASED AND SOC BASED CELL BALANCING TECHNIQUES [51], [52]

| technique | Measurement parameter | Pros | Cons | Application | Balancing accuracy |
|---|---|---|---|---|---|
| Voltage-based technique | Balancing based on monitoring and equalising cell voltages. | Easy to implement | • Balancing based solely on terminal voltage may not effectively equalise the SoC of each cell within the cluster.<br>• May only address cell imbalances at the beginning and end of the charging process, leading to incomplete balancing before charging concludes | Commonly used in low-cost consumer electronics where cell balancing is not that critical. | Poor |
| SoC-based technique | Balancing based on estimating and equalising cell SoC. Voltage differences between selected cells to generate balancing currents | More precise balancing | Requires more complex estimation algorithms. Higher implementation complexity. | Ideal for applications demanding high balancing precision. Found in advanced battery management systems for EVs, renewable energy storage, etc. | Good |

The choice of interconnecting cells in either series or parallel configuration depends on the specific application's requirements. The series cell connections are preferred in scenarios where achieving elevated voltage levels is a critical necessity [14].

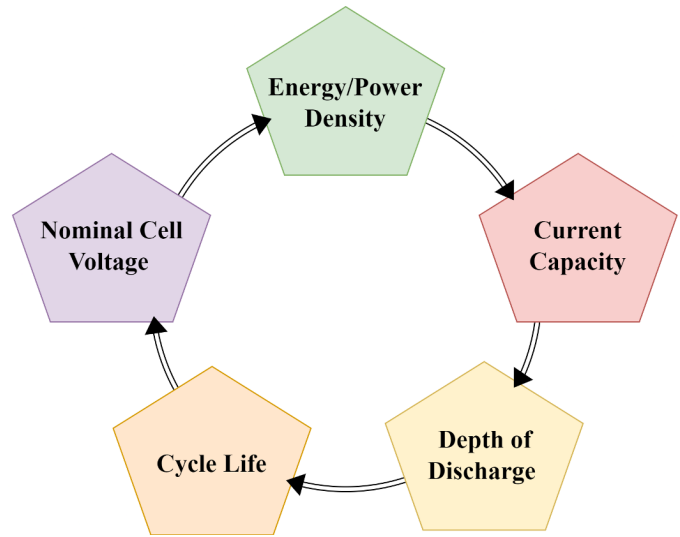

Fig. 4. Mutual dependence of battery key battery parameters [31]

In a series connection, the current through each cell remains constant; however, there are fluctuations in the voltage and SoC of each cell. The parallel cell connections are frequently implemented when applications demand increased capacity and current handling capabilities [53]. Specifically, parallel battery configurations utilise batteries with similar impedance [47].

In such connections, current flows through the branches, with each branch distributing based on the battery voltage and internal impedance to achieve equal branch voltages in the steady state [53], [54]. This phenomenon, known as the self-balancing effect, occurs momentarily [55]. Specifically, during idle periods without load current, energy transfers from the higher voltage batteries to the lower voltage batteries. However, due to impedance discrepancy, due to the ageing effect, the SoC differs between batteries, as shown in Fig.5, despite the presence of identical terminal voltages [56]. Further, during the non-idle periods when the batteries are charged or discharged by load currents, different current levels charge or discharge the battery packs [54]. These variations potentially result in long-lasting detrimental effects on the battery performance [57].

As shown in [54], a 20% variance in the cell internal resistance between two cells connected in parallel can result in around a 40% decrease in the cycle life. Thus, charge imbalance leads to circulating current, as shown in Fig. 6. This in turn causes increased energy loss, shortened battery life, and fire hazards [48]. Also, during the ageing process, inconsistencies are more pronounced as the impedance of battery cells continuously changes [47].

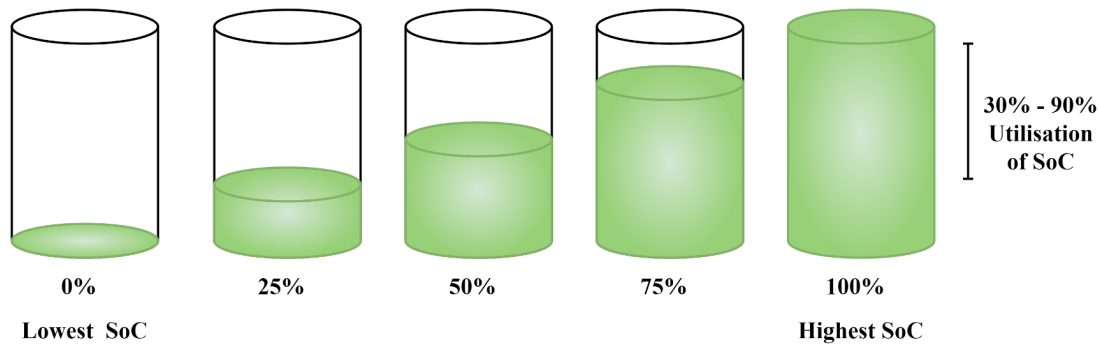

Fig. 5. Current circulation due to varying SoC levels among parallel connected batteries

For cell balancing, there are multiple techniques essential in applications ranging from EVs and renewable energy storage to portable electronics [58], [59]. Also, the cell balancing techniques implement various mechanisms to redistribute energy among the cells. The primary step in the process involves the design of an equalisation structure. This is the arrangement or mechanism that ensures that all the cells in a battery cluster have the same SoC or voltage level. This structure typically involves resistors, capacitors, inductors, or active electronic circuits facilitating energy transfer between the cells [60]. Based on this aspect, cell balancing techniques are broadly classified into active or passive techniques as depicted in Fig. 7 and detailed next.

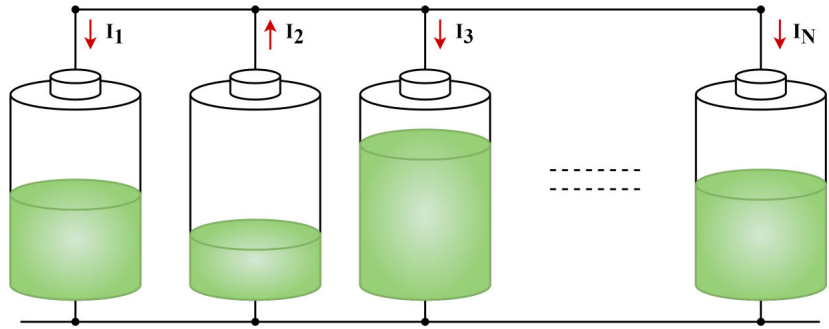

Fig. 6. Varied SoC levels among parallel connected batteries

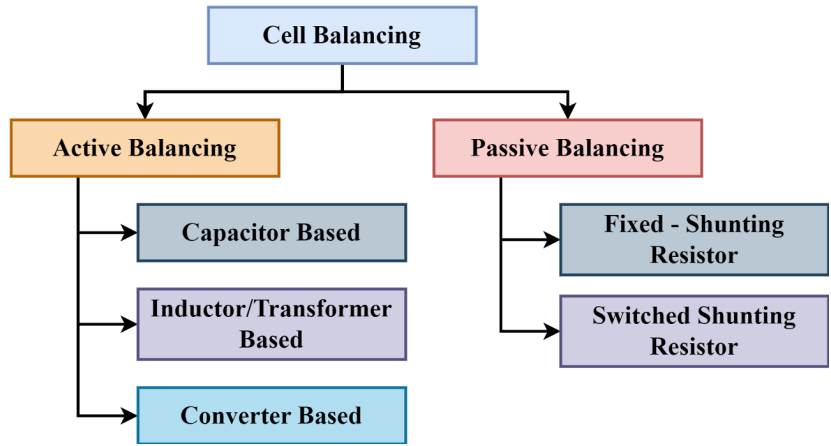

Fig. 7. Active and Passive cell balancing techniques

## A. Passive cell balancing

In the passive cell balancing approach, surplus energy in the high voltage cell is dissipated through a resistor until its voltage aligns with that of a weaker cell [59].

This technique is also known as dissipative cell balancing and is shown in Fig. 8. Additionally, this approach is easy to deploy and economically viable. It is used for small cells, i.e., requiring small power [29], with a balance current smaller than 10mA per Ah capacity of the cell [61], [62]. The Fixed Shunt Resistor and the Switched Shunt Resistor techniques are the categories of passive cell balancing [63]. They are compared in Table. V.

TABLE V
CLASSIFICATION OF PASSIVE CELL BALANCING TECHNIQUES [61]–[66]

| Type | Merits | Demerits | Applications |
| --- | --- | --- | --- |
| Fixed shunt resistor | • Low cost<br>• Easy to implement. | • Long-lasting energy losses.<br>• No controlled operation. | Appropriate for low-power applications utilising nickel and lead-acid batteries. |
| Switched Shunt resistor | • Simple controller<br>• Easy to implement. | • High currents in the balancing process lead to increased energy dissipation.<br>• Reduced balancing speed<br>• Involves the deployment of a greater number of switches.<br>• Best suited for use exclusively during charging. | • Suitable for Lithium-ion battery.<br>• Low power applications. |

### 1) Fixed Shunt Resistor

In the fixed shunt resistor approach, shunt connections of fixed resistors are recommended for each cell within the battery bank, as shown in Fig.9. The resistors are calibrated based on the desired balancing current [62] and are finely adjusted to constrain the cell voltages [64]. However, overcharging can happen because of the continuous current flowing through the cells arranged in series. In such cases, if the fixed shunt resistor is not adjusted correctly or there is a mismatch in the cell capacities, few cells may reach the full charge before others. As a result, the fully charged cells will continue to receive current, leading to overcharging.

Further, the popular LIBs require careful operation within specific safety parameters influenced by charge rate, temperature, and voltage range. In such batteries, voltage must be meticulously monitored and tightly controlled within the typical range of 4.1 to 4.3 volts per cell, as the threshold voltage for cell breakdown closely approaches the fully charged cell voltage [34].

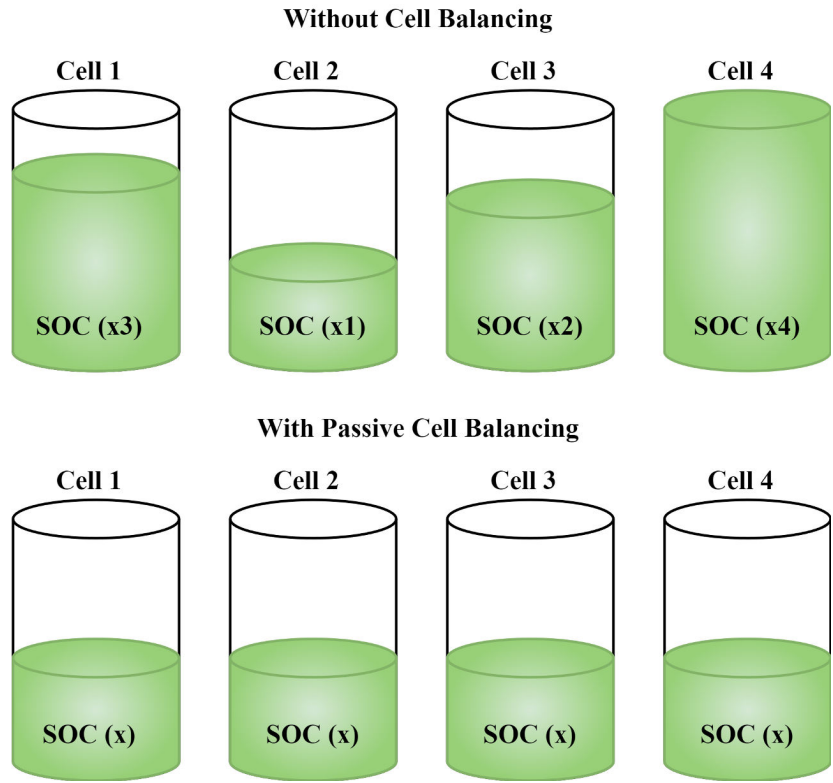

Fig. 8. Comparison between non-balancing and passive cell balancing approaches [62]

Deviating from these ranges can cause rapid deterioration of the battery performance and pose safety risks [67]. Ensuring this at all times in the case of a fixed shunt resistor is a challenge. Hence, this technique poses a risk of overcharging, especially in the LIBs [64]. Overcharging can result in cell degradation, reduced battery lifespan, and potentially hazardous conditions such as thermal runaway or cell rupture. Hence, meticulous monitoring and control of the charging operation are imperative to avoid overcharging and guarantee the safe functioning of the battery system. Studies in literature have shown that this technique is favourable for nickel and lead-acid batteries, which can endure overcharging without any adverse effects [62].

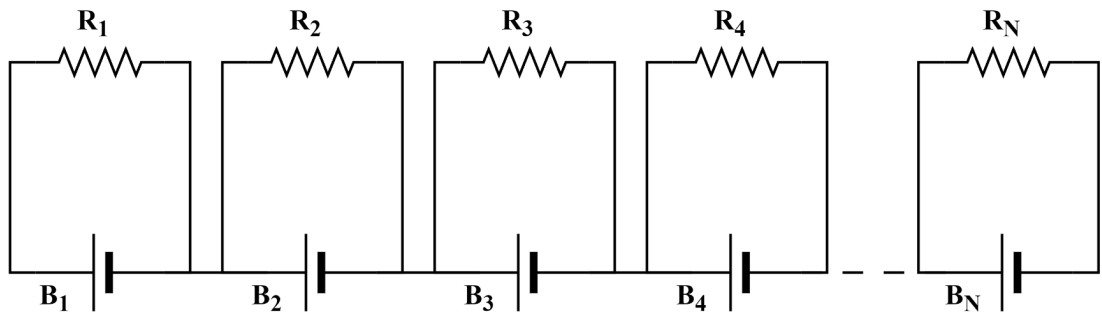

Fig. 9. Structure of Fixed Shunt Resistor [65]

*2) Switched Shunt Resistor*

The switched shunt resistor cell balancing technique, depicted in Fig. 10, involves connecting the resistors in parallel with each cell in a battery pack through the controlled switches. Unlike the fixed shunt resistors which remain continuously connected, the switched shunt resistors are turned on and off to balance the cells. It is also known as charge shutting and mitigates the voltage deviations between the cells by discharging surplus energy exclusively from the cells with higher voltage using the controlled switch on/off operations through a resistor [62], [65].

For LIBs, switched shunt resistors offer several advantages. They provide a more dynamic and flexible approach to cell balancing, allowing for precise control over the balancing process. This technique can also efficiently equalise the charge among cells without dissipating excess energy as heat by selectively activating the shunt resistors based on the cell voltage levels. Thus, this technique can prevent the overcharging of individual cells and extend the overall lifespan of the battery pack.

Additionally, the switched shunt resistors also minimise energy losses during the balancing process thereby, enhancing the battery system's efficiency [62].

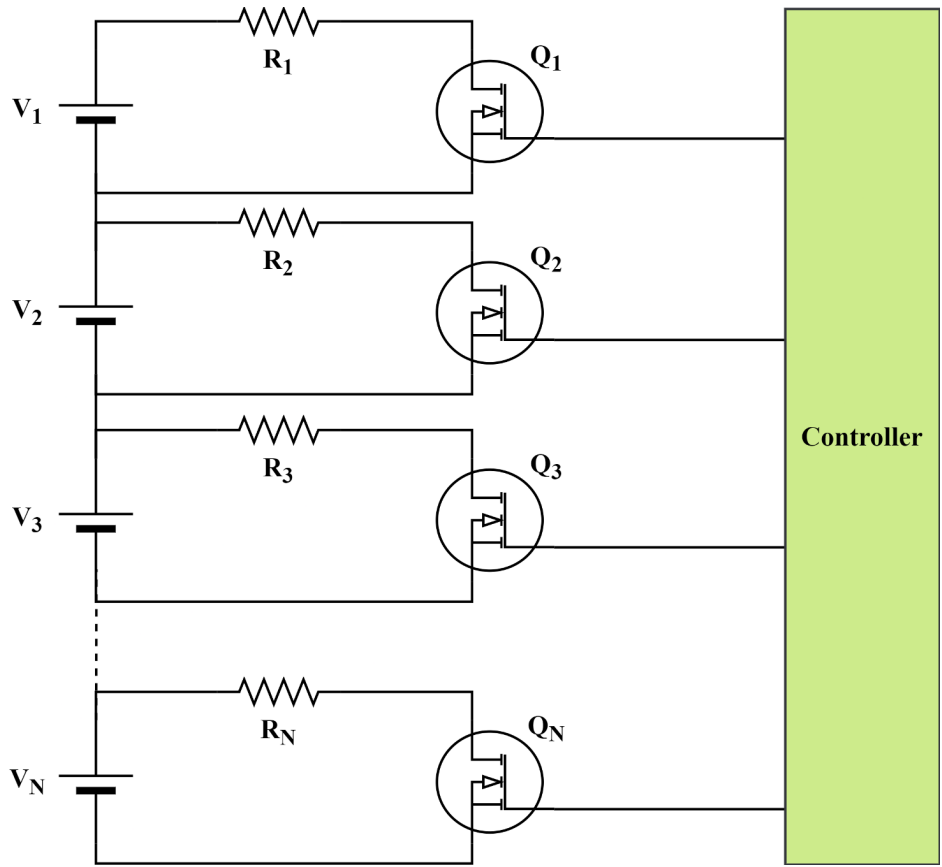

Fig. 10. Structure of Switched Shunt Resistor [62]

### B. Active cell balancing

The effect of this technique is shown in Fig. 11. Herein, the energy storage elements such as capacitors and indicators are used to store the excess energy.

Any surplus energy is redistributed to cells with lower voltage within the battery cluster to equalise the cell voltages [10], [68]. This technique offers advantages such as faster balancing, higher efficiency, reduced heat generation, etc. The majority of the active cell equalisation techniques discussed in the literature use one of the equalisation structures listed in Table. VI. Specifically, the table comprehends every approach's pros and cons, and tabulates the application of each approach. It must be noted that choosing the most suitable balancing approach depends on user's specific applications, battery pack size, balancing requirements, and cost constraints [4], [14], [69]. A detailed categorisation of the active cell balancing techniques is determined by the components employed to achieve cell balancing. As a result, these techniques are classified into capacitor-based, inductor-based, and converter-based techniques. A comprehensive comparison of these techniques is provided in the Table. VII.

### C. Cell balancing through Charging

Another alternative technique, known as continuous charging, involves charging the cells until they are all balanced to a certain level. This ensures the controlled overcharging of cells, resulting in the full charge of the higher capacity cells. This technique applies to Pb-acid and Ni-based batteries as they can tolerate overcharge to certain degree without substantial damage.

However, careful implementation of this technique is necessary, as excessive overcharge can result in cell overheating and ultimately early drying of the electrolytes. Despite its simplicity and cost effectiveness, this technique has drawbacks, including lower efficiency and lengthy time needed to achieve cell balance [34].

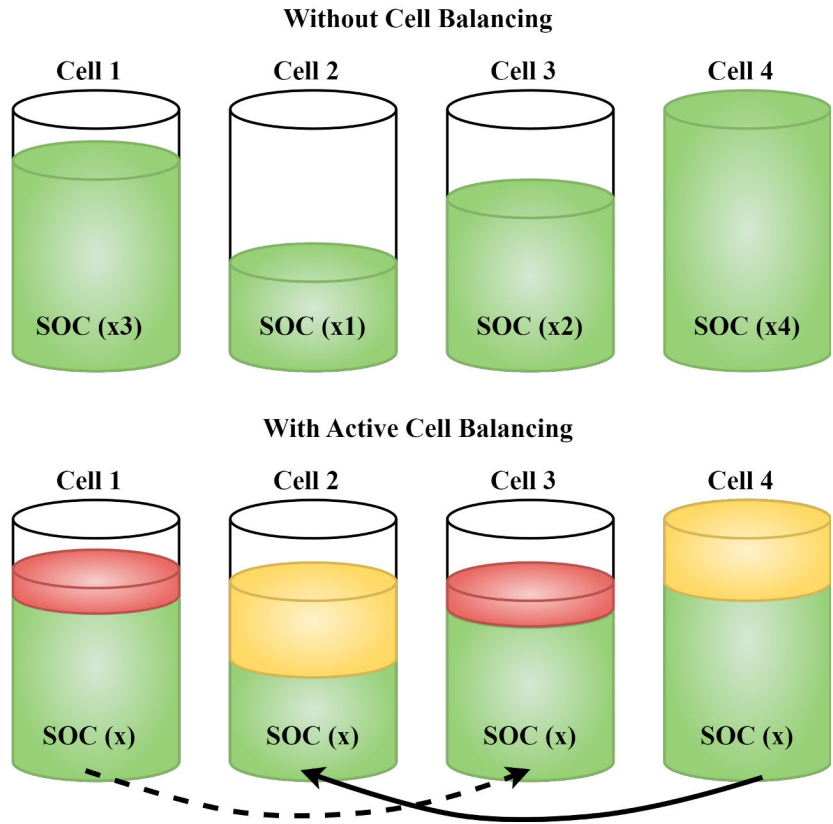

Fig. 11. Comparison between non-balancing and active cell balancing approaches [62]

## III. RECENT ADVANCES ON CELL BALANCING

A fixed resistor is conventionally employed to attain passive cell balancing. The duration of the balancing process using this technique is shorter when cell imbalance is minimal. However, if there is a significant imbalance, it can result in a longer balancing time. To tackle this problem, a simple, low-cost passive balancing circuit is proposed in [71], with an option for selecting the desired resistor. This approach can be applied to both new and aged cells, depending on the need to balance current during either slow or fast charging scenarios. A passive equalisation circuit consisting of a resistor and a switch is proposed in [72]. A Pulse-Width Modulation (PWM) technique is utilised to regulate the operation of the switch. The design procedure and recommendations for the high power balancing circuit are detailed emphasising the operation region of the MOSFET, the selection of switching frequency, etc. Excessive heat dissipation issues of passive cell balancing techniques due to the balancing resistors and high balancing time are addressed in [73]. The algorithm formulated therein selects an appropriate resistor to increase the balancing speed of the system with a reduced power loss. The algorithm is implemented via a neural network considering real-time input parameters such as charging time, charger utilisation, charging frequency, charging temperature, battery chemistry, thermal effect, and probability of imbalance.

Comparative analysis with traditional approaches reveals that the recommended algorithm significantly increases the cell balancing speed. The study in [74] emphasises the significance of tackling problems associated with variations in cell parameters and different operational conditions in series-connected cells of EV batteries. The addressed issues include imbalances in SoC and temperature of the cells. The authors introduce a redundant smart battery architecture where cell modules connected in series are provided with an insertion/bypass circuit. The study details the design of a controller based on K-nearest ML algorithm, which is designed for cell insertion/bypass to achieve instantaneous balancing of SoC and temperature.

A current mode controller is proposed in [5] for a switched-inductor based battery balancing topology to equalise voltage or SoC of the battery cells. The control scheme uses a buck-boost converter and overcomes the limitations of traditional techniques, where the duty cycle is fixed. It is observed that conventional techniques lead to more losses and hence reduced efficiency. The authors also present a technique to maintain the inductor current in an acceptable safe value and hence the balancing current, despite the latter being a function of many uncontrollable parameters. Thus, the operation at maximum efficiency with reduced power losses is ensured. Also, the authors present an open circuit voltage balancing for accurate SoC measurements. AC2AC battery equaliser utilises a half-bridge LC converter operating at resonance is proposed in [69]. Flexible balancing modes, including cell-to-cell, cell-to-multi-cell, multi-cell-to-cell, and multi-cell-to-multi-cell are discussed. This is shown to be applicable irrespective of the number or position of the target cells. The different components include a half-bridge, two resonant capacitors, a resonant inductor, and a transformer winding per cell. The study presents advantages such as improved speed of energy balancing, and reduced usage of passive components and complexity. Finally, the proposed technique is shown to be suitable for recycled batteries. The limitations of various non-dissipative balancing circuits, limited balancing speed, low balancing efficiency, complex circuitry etc., are detailed in [4]. The authors introduce an improved SC-based balancing circuit and three-resonant-state LC unit for AC2AC and DC2C, respectively. A two-mode balancing circuit combining these two techniques is presented to surpass the limitations listed above. Additionally, the article presents a study examining how circuit parameters affect the speed and efficiency of the balancing process.

TABLE VI
CLASSIFICATION OF EQUALISATION STRUCTURES [4], [14], [44], [60], [69], [70]

| Equalisation Structure | Energy transfer | Pros | Cons | Applications | Balancing speed |
|---|---|---|---|---|---|
| Adjacent cell to cell | Between two adjacent cells | Suitable for small- to medium-sized battery packs. | • Limited in addressing imbalances between non-adjacent cells.<br>• Efficiency issue due to energy losses during cell-to-cell transfer.<br>• Increased complexity and expense of the balancing circuit. | Can be incorporated within individual cells or compact modules | Decreases with the increase in cell number |
| Direct cell to cell (DC2C) | Between arbitrary two cells regardless of their positions. | Use of a single equaliser, Provides a more direct and precise approach to balancing. | • May lead to increased wear and tear on components.<br>• Difficult to incorporate in large batteries.<br>• Works efficiently only if designed well.<br>• Balancing imbalanced cells necessitates a longer duration to ensure thorough adjustment | Effective for small to medium-sized battery packs. | Lower balancing speed for a long cell string. |
| Cell to pack | From an individual cell directly to the high voltage pack. | • Balances the entire pack collectively, addressing non-adjacent cell imbalances.<br>• Can be more efficient due to collective balancing. | • May not provide individual cell-level balancing.<br>• Certain balanced cells may experience unintended charging or discharging, leading to potential issues.<br>• The non-targeted cells must temporarily halt their operations until the balancing process of the specific cells is completed.<br>• Adaptability for the commercial electric vehicle (EV) market is challenging.<br>• Each circuit demands either a significant number of inductors or transformers, or a large number of switches, leading to increased cost.<br>• Reduced conversion efficiency and increased voltage stress on power switches. | • Suitable for larger battery packs with multiple cells.<br>• Very ineffective and time consuming when only a specific cell is weak | High. |
| Pack to cell | From the high voltage pack to the target cell. | • Allows for modular and scalable design.<br>• Suitable for various pack configuration. | • Requires sophisticated control algorithms and communication between cells and the central BMS.<br>• May not address imbalances at the cell level.<br>• Few balanced cells may undergo unintended charging or discharging.<br>• The completion of balancing in the target cells necessitates the waiting of other cells until they finish.<br>• Adaptability for the commercial EV market is challenging. Each circuit demands either a considerable number of inductors or transformers, or a greater number of switches, leading to increased cost | Effective for balancing a group of cells within a pack. | High |
| Cell to String | From higher voltage cell to the cell string | Improved balancing speed | Implementing direct energy transfer between imbalanced cells is not feasible. | Effective for balancing a group of cells within a pack. | Moderate |
| String to Cell | From the cell string to the lower voltage cell | • Simple, reduced complexity<br>• Improved equalisation speed | Implementing direct energy transfer between imbalanced cells is not feasible. | Effective for balancing a group of cells within a pack. | Moderate |
| Any cell to any cell (AC2AC) | From all higher-voltage cells to all lower-voltage cells. | • Simple circuit with small size, reduced cost, and lower voltage stress on components.<br>• Offers flexibility for various imbalances. | • Complex and longer balancing paths.<br>• Lower efficiency for the multi-cell string. | Suitable when individual cells or groups of cells are to be balanced. | • Usually Lower.<br>• Higher when multi-winding transformer is used. |
| Multi-cell to Multi-cell | From the higher voltage sub-string (a sub-string consists of multiple cells in series) to the lower voltage sub-string. | Simplified balancing process. | • Implementing direct energy transfer between imbalanced cells is not feasible.<br>• Lower balancing efficiency.<br>• Highly dependent on the particular converter used to build the equaliser. | • Suitable for applications with numerous cells.<br>• Effective in high-capacity battery packs | Moderate |

In [11], the authors extend the range of an EV by improving the performance of its battery. A combination of real-time realisable active cell balancing technique with long-term ageing data is used to demonstrate that active cell balancing significantly extends the life span of a battery cluster. The performance of the proposed technique is shown to be better than existing techniques with regard to reducing computation time and enhancing reliability. The results also demonstrate a substantial extension of the battery pack's End-of-Life (EoL) by 10%. A technology to achieve SoC balance in a three-phase battery energy storage system using three cascaded hybrid modular multi-level converters is discussed in [8]. The article also introduces a Battery Energy Storage System (BESS) topology that simplifies the design by employing a single branch instead of three in a three-phase battery energy storage system. The strategy is designed to work with or without the idle cells/modules. Thus, this eliminates the necessity for SoC balancing among the branches, as in conventional topologies. In contrast to the conventional topologies, these idle cells/modules are shown to be able to serve as redundancy or be removed from the BESS. A novel algorithm is detailed to reduce the duration required to activate the cells/modules. Also, the control complexity, size, losses, and BESS total cost are reduced. The study in [75] tackles the limitations associated with conventional techniques for extending the durability of batteries, where the full utilisation of the battery's available capacity is often hindered. However, this can be achieved by the use of intelligent BMS. A long-term test is conducted by exposing two batteries to the same loads. One battery is connected to an active BMS and the other to a traditional passive balancing BMS. It is shown that an active BMS can enhance life of the battery.

The passive balancing technique is inherently dissipative, yet it offers advantages such as cost-effectiveness and easy implementation. The appropriate selection of the balancing resistors offers a feasible solution to mitigate its limitations. In [73], Neural Network algorithms are applied considering various real-time parameters that influence the battery working. This is ensured to optimise parameters such as balancing time, power loss, and temperature rise. It is shown that the proposed technology overcomes the disadvantages of the prevailing passive balancing systems. The authors in [76] present a passive balancing BMS technique that has the advantage of a power resistor and MOSFET internal resistance as a balancing resistor. The recommended technique conserves space on BMS hardware compared to a power resistor with a larger balancing current capability. The obtained results show that when the technique is applied to a 1P15S battery pack with a 200 Ah LiFePO4 battery capacity, the balancing current of the battery is determined by the MOSFET's internal resistance value.

TABLE VII
CLASSIFICATION OF ACTIVE CELL BALANCING TECHNIQUES [4], [14], [44], [60], [69], [70]

| | Types | Advantages | Disadvantages | Number of Elements | Equalisation Structure |
|---|---|---|---|---|---|
| Capacitor-based | Switched capacitor technique | Simple structure and no complex control algorithm [60] | High equalisation time [60] Lowers the equalisation current and the overall efficiency of the battery is reduced. [68] | n-1 capacitors 2n switches | adjacent cell to cell |
| Capacitor based | Single switched capacitor | Efficient, cost-efficient, low complexity, low switch voltage stress, simple controller design, suitable for low and high power applications | Difficult modularity; high switch voltage stress; highly complex and Long equalisation time | 1 capacitor n+5 switches | Adjacent cell to cell, cell to pack to cell. |
| Capacitor-based | Double Tiered Capacitor | Lower balancing capacity currents, high power applications, easily modularised, reduced balancing time. | Relatively lower speed; higher switch current stress. | n capacitor 2n switches | Adjacent cell to cell, cell to pack to cell. |
| Capacitor based | Modularisation switched Capacitor | Adequate balancing speed and higher efficiency | Expensive. | n capacitors, n+2 switches | Adjacent cell to cell, cell to pack to cell |
| Transformer based | Single transformer | Reduced component count | Complex control, high cost | n+1 switches, 1 transformer | pack to cell, cell to pack, DC2C |
| Transformer based | Multi transformer | Higher balancing speed and efficiency | High magnetic losses, Expensive | n switches, Single transformer with multiple secondary windings | DC2C, cell to pack, pack to cell. |
| Inductor-based | Single inductor | Satisfactory balancing speed High efficiency. | Complex control, filtering capacitors needed for high switching frequency | 1 inductors, 2n switches, 2n diodes | pack to cell, cell to pack, DC2C, cell to pack to cell. |
| Inductor based | Multi Inductor-based balancing | Higher balancing speed and efficiency | Higher switch current stress | n − 1 inductors, 2n-2 switches | pack to cell, cell to pack, DC2C |
| Converter based | Buck-boost converter | Simple, negligible power loss, high conversion efficiency | Low balancing speed, Control complexity. | n converters | pack to cell, cell to pack, DC2C |
| Converter-based | Bidirectional Cuk converter | Less expensive | Large size of circuit topology | n converters | pack to cell, cell to pack, DC2C, cell to cell. |
| Converter-based | Quasi-Resonant converter | Less switching losses, high energy efficiency | Execution difficulty | n-1 converters | pack to cell, cell to pack, DC2C, cell to cell |
| Converter-based | Fly-back converter | Utilises a simple circuit with minimal number of circuit components | Transformer is one essential component | n+4 switches 1 converter | pack to cell, cell to pack, DC2C, cell to pack to cell. |
| Converter-based | Cascade full-bridge (H-Bridge) multilevel converter | Suitable for high power applications and enhanced equalisation speed, and modular structure. | High voltage/ current stresses, large size, and expensive. | (2 n + 1) H-bridges | cell to pack, DC2C, pack to cell |
| Converter-based | Dual active bridge converter (DAB) | Suitable for a battery cluster with a increased number of cells and provides higher balancing speed. | Bulky size and high cost. | n+3 switches,1 converter | pack to cell, cell to pack, DC2C, cell to pack to cell, cell to cell |
| | Forward converter | Complicated circuit | Reduced balancing efficiency | n+3 switches, 1 converter | pack to cell, cell to pack, DC2C, cell |

Charge imbalances are typically resolved by either supplying higher current to undercharged batteries or transferring charge between batteries to equalise levels. The authors in [77] proposed a technique combining the two previously mentioned, i.e., battery reconfiguration and charge transfer, to further improve cell balancing. Battery reconfiguration algorithms are proposed to reduce battery charge equalisation time by battery cell/module reconfiguration. The effect of battery cell/module configuration on the speed of battery charge equalisation is discussed.

The circuit proposed in [58] provides a solution for both, active and passive balancing, incorporating a high precision 12-bit Analog-to-Digital Converter (ADC) with an accuracy level of ±7mV. It incorporates balancing switches within the circuit design, effectively minimising the need for external components. This approach not only reduces the overall complexity of the balancing circuit, but it also addresses cost-related concerns as it optimally utilises a minimal number of external components. The performance of the fly-back converter for cell balancing depends highly on the coupling coefficient of the transformer. A lower value of the coupling coefficient may lead to transferring energy to non-target cells, thus resulting in cell imbalance and reduced system efficiency. To circumvent this problem, a new switching pattern based on fly-back converter is proposed in [10]. The switching scheme utilises non-participating cells to control voltage across each winding, ensuring a higher energy transfer ratio, independent of the coupling coefficient. Additionally, a novel technique has been introduced to accelerate cell balancing by minimising the energy transfer path in certain cell conditions.

The authors in [78] provide a comprehensive overview of active cell equalisation circuits for BMS in EVs, highlighting their advantages and use cases with specifications. Further, the study also discusses different converter topologies with high energy efficiency for active cell balancing. Overall, the study provides a comprehensive overview of active cell equalisation circuits and compares converters in terms of components and performance. Among the different topologies used for active cell balancing, transformer-based techniques are commonly employed. The use of a multi-winding transformer-based equaliser requires additional demagnetising circuits, many switches, and sophisticated control logic, resulting in bulk size and a costly balancing system. As a solution, authors in [43] proposed a low-cost cell balancing technique for retired series-connected battery string using a multi-winding transformer. The inductor of a DC–DC converter is used with the balance system by utilising its current ripple. The technique operates with an efficiency of 94.1% at rated 28 W input power. Further, the proposed technique does not require additional switches and a controller. It is shown that a cost reduction of 60% is achieved when compared to traditional transformer-based approaches.

When cells are arranged in parallel and exhibit differences in terminal voltage, there is a circulating current flow and variations in SoC among the cells. Considering this aspect, an active cell equalisation approach for parallel connected LiFePO4 battery cells is proposed in [56]. The proposed technique utilises an adaptive Back Propagation Neural Network (BPNN) for SoC estimation, with an estimation error of about 1.15%. The proposed approach in [79] combines the benefits of both active and passive balancing techniques. Simultaneously, it also addresses the limitations of passive techniques, such as the direct connection of cell terminals to a balancing board, which can pose safety and assembly challenges, especially in EVs. The technique allows cell voltage monitoring and active cell balancing. It also provides complete isolation with low-voltage wiring, making it well-suited for high-voltage applications with rigid safety requirements, including automotive and industrial EVs. The approach is simple and uses only one isolation transformer for each cell. The balancing efficiency is reported to be 40% to 60%, but the authors state that there is room for further efficiency improvement. In the case of various non-dissipative balancing techniques, unnecessary charge transfers to the cells affect the battery life span. To address this issue, a novel technology that employs a multi-winding input and output fly-back converter is proposed in [80] to enhance the balancing accuracy. The technique aids in the balancing of multiple cell pairs flexibly. The initial balancing sequence is determined by the cell SoC, thus simplifying the control logic. The performance of this technique is assessed with the standard fly-back converter-based topologies in terms of time required for balancing and system efficiency. It is shown that the proposed technique is approximately 2-3 times faster in balancing and improves efficiency by 0.4%.

Notably, certain parameters of the battery model exhibit significant fluctuations even under consistent temperature and discharge current rate. In such situations, model-based solutions are not effective for SoC estimation. Hence, a novel and practical technique for achieving active cell balancing is proposed in [81]. The study discusses that consistent cell balancing can be achieved even by using simple SoC estimation techniques like OCV or even Coulomb counting. The technique trades-off between convergence time and safety, and the results indicate a shorter charging period and increased battery efficiency. Deploying high-frequency switched MOSFETs necessitates a complicated gate-driver leading to higher expenses, potentially exceeding the cost of the MOSFETs themselves. This issue is addressed in [82] by proposing a cell-to-cell voltage equaliser encompassing low-frequency selection switches based on a capacitive level-shifted Cuk converter. The design excludes isolation transformers and diodes resulting in high efficiency. The utilisation of low frequency switches with a simple driver circuits is also recommended to reduce component count and cost. The inclusion of bipolar voltage buses is shown to reduce the switch count. A novel, hardware implementable fast cell balancing technique is proposed in [14] for maximising the EV driving range. The presented low voltage-to-cell battery balancing technique offers the advantage of reduced component number compared to existing fast equalisation topologies. An efficiency of approximately 70% is achieved by incorporating a fly-back converter. The proposed circuit also enables rapid balancing during driving by facilitating concurrent energy transfer between non-adjacent cells throughout the pack. Also, through system-level vehicle modelling, an increase in driving range from 1.8% to 20.1% is demonstrated. This improvement is observed when the proposed technique is compared with passive balancing, which does not engage in cell charging while the EV is in motion.

The architecture proposed in [59] uses two non-isolated DC-DC converters to achieve active cell performing using MOSFET switch matrix. The balancing of an arbitrary cell at a high current is discussed. The results demonstrate a balancing cell accuracy of 92.5%. A three-stage charging technique is proposed for Lithium battery in [3]. The process involves pre-balancing stage, a constant current charging stage, and a constant voltage charging stage. Cell balancing is achieved in the last two stages, and the study is conducted on six series and two parallel battery packs. The balancing time is observed to be 3600 seconds. A comparative study of dissipative, capacitive, and runtime cell balancing techniques in terms of efficiency, cell balancing speed, complexity, and cost, is presented in [83]. The runtime balancing technique is fast and efficient; however, it is complex and expensive. The technique employs a lower power converter to boost the voltage to the required value. It is shown that the power converter is not required for charging/discharging control, and it can be applied to pre-used batteries.

The problems of a long balancing time and a large number of external components related to non-centralised techniques are addressed in [84]. A centralised charge equalisation system encompassing Integrated Cascaded Bidirectional (ICB) converter is presented. The advantages of the technology include reduced use of power switches, significant improvement in the transformer efficiency, and reduced switching loss. The proposed technology is experimentally validated, and an improved balancing speed is achieved. The challenges associated with the traditional cell-to-pack-to-cell equalisation system, including issues with compatibility for large-scale battery packs, repair and maintenance complexities, and the oversight of cell current constraints, are discussed in [85]. To address the limitations, a modular technique is introduced, which combines cell-to-module-to-cell and module-level equalisers to achieve SoC equalisation. The article further elaborates on the application of a two-layer Model Predictive Control (MPC) strategy and the utilisation of the Lyapunov Stability Theorem for addressing these challenges.

The function of traditional active cell equalisation techniques depends on the position of the inductor and the number of cells in a module. The authors in [61] introduce a novel active balancing circuit for N cells connected in series using a buck-boost converter and low-frequency bidirectional switches to control the state of inductors, connection, disconnection, and short-circuit. The advantages offered by the proposed technique are improvements in terms of output power, flexible cell group formation, versatile operating modes, and low power loss. Further, the proposed technique is also shown to improve the equalisation speed. The existing limitations associated with active cell balancing techniques such as low efficiency, large volume, reliability issues, and extended cell voltage equalisation time are addressed in [68]. The authors introduce a novel approach viz., closed-loop switched-capacitor structure, to achieve the shortest path for voltage equalisation between individual cells within a battery pack. An experimental validation of the technique is conducted, and it is demonstrated that it is capable of balancing the voltage of any cell with any other cell in the pack. The proposed converter structure exhibits an efficiency of up to 95%. The technique is evaluated considering various switching frequency conditions and is shown to provide the remarkable advantage of equalising cell voltages regardless of their initial voltage levels. An active cell balancing approach is proposed for low voltage applications in [86]. The use of a non-isolated DC-DC converter and a low-speed switching matrix without the use of external energy tanks such as super-capacitor is proposed. The study is conducted considering eight randomly distributed battery cells, and a balancing efficiency of 90% is observed.

A modular equalisation system for LIB modules is introduced in [87]. The proposed system offers dual equalisation modes (cell and module) and implements the Dual Phase-Shift (DPS) control technique. Inter-module equalisation is facilitated through adjacent modules' Capacitively Isolated Dual Active Bridge (CIDAB) converters connected via an LC tank. The system is scalable, maintaining intra-module equaliser consistency. Experiments with two LIB modules validate the system's effectiveness, making it a flexible and promising solution for efficient battery balancing. An approach for equalising parallel battery configurations through dynamic resistance adjustment, aimed at enhancing equalisation performance, optimising battery capacity, and reducing dissipation losses is proposed in [47]. Using the SoC status of battery cells as a reference, the switches are controlled to alter the impedance of parallel branches, thereby regulating branch currents. To validate the proposed technology, simulations are performed using relevant data from four 18650 Li-ion battery cells, each with 3.7V/2.6Ah capacity. Authors in [88] propose a new approach that (i) implements dynamic capacitance modulation to balance module voltages during idle periods, and (ii) efficiently regulates branch currents during active periods. The impact of the recommended technology is validated through hardware-in-the-loop tests involving four parallel-connected battery modules. The results indicate SoC equalisation within 1% during idle periods and the suppression of inconsistencies during active periods. The study in [55] introduces an equaliser and control strategy designed for parallel-connected battery packs. Utilising one bidirectional converter and a switch-matrix, the proposed approach ensures equalisation of SoC levels across battery packs while addressing the inrush current problem during hot-swap processes. Test results demonstrate the effective performance in idle, charging, and discharging modes, achieving SoC equalisation within a 0.5% tolerance without overloading conditions.

The detailed survey is summarised in two parts in (i) Table. VIII, which presents the classification of various active cell balancing techniques, outlining their respective advantages, disadvantages, number of elements, and preferred equalisation structures; and (ii) Table. IX, which presents a summary of passive cell balancing techniques. Overall, the detailed survey conducted in this section allows answering the specific research questions listed in Section I. The literature review thoroughly examines the existing literature and aids in addressing the diverse aspects of cell balancing, including the factors contributing to cell imbalance. Further, the survey also explores various circuit topologies suitable for both, series and parallel connected cells, detailing their advantages, drawbacks, and component compositions. Furthermore, the survey discusses techniques implemented to prevent overcharging and discharging in series connections. It is found that the primary aim is to alleviate cell imbalances, enhance balancing accuracy, and reduce balancing time. Additionally, the review highlights the importance of cell balancing across commonly used battery types.

## IV. Cell Balancing in Transportation Sector

Within the complex energy landscape, renewable resources are increasingly asserting their presence. While integrating them within the existing infrastructure is feasible, infrastructure modifications are necessary. Further, the growing use of renewable sources presents numerous challenges owing to intermittent generation. To effectively address this issue, the use of storage devices such as batteries is essential [34], mandating the implementation of cell balancing. Notably, the most popular and well-established use case/application of cell balancing is in the transportation sector which significantly contributes to greenhouse gas emissions, accounting for approximately 23% of carbon dioxide emissions [93]. The need to implement cell balancing in batteries for the transport sector is illustrated in Fig. 12. Notice that the key task in this sector is to identify the availability of a specific battery type to be used. Once completed, it will be required to (i) implement BMS, which will incur additional costs; and (ii) identify techniques with appropriate methods for implementation. In this regard, research must be directed towards formulating complex techniques to identify the appropriate solutions and simultaneously provision the desired safety needs.

In the transportation sector, batteries are playing a pivotal role in revolutionising transportation, particularly in EVs and hybrid EVs (HEVs) development [94], [95]. In EVs, batteries serve as the primary energy storage solution, powering the

TABLE VIII
ACTIVE CELL BALANCING TECHNIQUES: A SURVEY-BASED COMPARATIVE ANALYSIS

| Reference | Switch | R | L | C | Diode | MOS | Transformer | Switch f (Hz) | Balancing I (A) | Drawback | Future scope as indicated |
|---|---|---|---|---|---|---|---|---|---|---|---|
| [82] | n+2 DPDT, 2 SPST | 0 | 2 | 2 | 2 | n+1 | 0 | - | - | - | |
| [14] | n+3m | 0 | 0 | 2m | 0 | 2(n+3) | m | 100k | 5 and 10 | high hardware requirement | - |
| [43] | - | 0 | 4 | 0 | 4 | 0 | 0 | - | 2.1 | - | |
| Non synchronous design [59] | 2n-2 | 0 | 2 | 2 | 2 | 4 | - | - | 2.5 | Use of high power rating of the converters | Exploring the use of lower-rated converters |
| Synchronous design [59] | 2n-2 | 0 | 2 | 2 | 0 | 4 | - | - | 2.5 | Suitable for a battery with 7 or 8 cells. | Explore the possibility of extending the work for a battery with more than eight cells |
| [89] | 6n | 0 | 0 | 1 | 0 | - | n | 1M | 2 | - | Realisation with reduced number of switches to 2/cell |
| [15] | m+1/(m*n) | 0 | 0 | n | 0 | n+1 | N+1 windings | - | - | - | - |
| [90] | - | 0 | 1 | 1 | 0 | 4n | 0 | - | - | - | |
| [91] | - | - | n | n | 0 | 4n-2 | 0 | 4.1k | - | - | - |
| [61] | - | - | n-1 | 0 | 0 | 5n-4 | 0 | 10k | - | Hard to achieve equalisation between nonadjacent cells. Large circuit size | |
| [92] | HF:2n-2 LF:n-1 | - | 0 | - | 0 | - | 0 | 10k | - | - | - |
| [86] | n+1 | 0 | 1 | 0 | 0 | 4 | 0 | - | 5 | High-power rating of the converter and its associated losses. | - |

*n- number of cells, m-number of modules, R-resistor, L-inductor, C-capacitor

TABLE IX
PASSIVE CELL BALANCING TECHNIQUES: A SURVEY-BASED COMPARATIVE ANALYSIS

| Reference | Other components | Balancing Switch | External R | Switch f (Hz) | Balancing current | Operating voltage of each cell of battery | Drawback | Scope | Hardware experimentation |
|---|---|---|---|---|---|---|---|---|---|
| [72] | - | 1 | 1 | 1, 50, 250 and 500 | 0.- | 4 to 4.2 V | Increased MOSFET temperature with a decrease in bleeding resistance | An appropriate design is necessary to find an optimal configuration of the power losses in the MOSFET | No |
| [58] | ADC-12 bit A synchronous voltage mode level shifting circuit | Internal | 0 | - | 100m | 4V | Increased system complexity for cells more than 7. No mention about balancing speed | - | Yes |
| [71] | - | 2 per cell | Dependent on the required balancing current | - | 100-400m | 4.2 and 2.5 V | - | - | No |

electric motors that propel the vehicle. This enables EVs to operate without traditional internal combustion engines which aids in significant minimisation of greenhouse gas emissions and air pollution [96]. The preferred characteristics of batteries for use in the EVs include [1], [34]

- Enhanced specific energy (Ah/weight), leading to extended All-Electric Range (AER) and reduced frequency of recharge cycles.
- Superior specific power, facilitating rapid acceleration in Plug-in HEVs (PHEVs), enabled by the battery's capacity to deliver high currents without compromising its longevity.
- Increased durability with many charge/discharge cycles and robust safety features inherent in the battery design, attributed to its elevated power capabilities.
- Eco-friendly composition, characterised by recyclability and minimal hazardous material presence, aligning with sustain-

ability goals.

Further, cell balancing is important for EVs as it helps extend the vehicle's driving ranges and ensures safe EV battery operation. In general, cell balancing in EVs comprises three different stages viz., (i) Idle mode, (ii) charging mode, and (iii) discharge mode. Active balancing during the idle mode is inefficient and results in major power losses in comparison to passive balancing [97]. Meanwhile, active balancing techniques utilised during the charging or discharging cycles are more effective, although they generate immense amounts of heat, leading to high costs. Therefore, active balancing is more applicable for high-power battery packs, whereas passive balancing can be implemented for low-power battery packs. Notably, active and passive balancing techniques can be improved using optimisation techniques based on machine learning [98]. These optimised balancing techniques will aid in reducing the drastic temperature variations caused by internal or external factors, as well as in minimising the power losses [99].

In EVs, battery cell imbalance greatly affects battery cell performance, which in turn has significant impact on EV overall performance. A key issue that reduces capacity is the early charge termination and over discharge. This leads to increase in the impedance and longer operational periods. Cell voltage drift due to imbalance within the battery system can lead to over-voltage exposure and premature degradation, resulting in reduced capacity and potential safety hazards such as explosions [100]. Additionally, the cell equalisation control unit of BMS terminates discharge to prevent the over discharge of the cells and any resulting damage if any cells reach low voltage threshold; however, this reduces the battery capacity. Hence, in general, cell based termination voltage is set to a lower value than pack based threshold divided by serial cells amount so that this difference allows for small imbalances [13]. Also, premature cell degradation occurs via exposure to overcharging in which case, SOC or total capacity imbalance is caused, wherein cells with higher resulting SOC are exposed to higher voltages. Also, low-capacity cells are affected by much higher voltages than other cells, while normal capacity cells have lower voltage than that attained during normal charging [101]. These aspects are currently being addressed in the commercially available EVs to improve their performance further.

Additionally, batteries are used in HEVs, which combine both, conventional internal combustion engines and electric propulsion systems. Specifically, in the HEVs, cells are wired in parallel thereby forming a block for satisfying requirements of high capacity, simultaneously ensuring that multiple cells are connected in series to provide a high voltage [102]. Also, since every cell is distinct due to manufacturing and chemical offset, cells in series include same current but different voltages. Therefore, during charging, capacity fade in cells may result in danger if a cell completely charges i.e., it will suffer from overcharging while the remaining cells reach the complete charge. Similarly, over discharge may occur on weakest cell, which will fail before others during discharging process. Also, when battery consists of multiple cells in series, it will be subject to a higher failure rate than any single cell, an intrinsic property of a series network. To minimise such an effect and extend battery life, an effective cell balancing mechanism is required. Such a mechanism must ensure SOC levels of individual cells in a battery pack are close to each other [103]. However, this will require charge or energy exchange among cells, which will result in complicated charge-discharge profiles in comparison to the conventional profiles. Overall, the balancing techniques in HEVs will depend on determining SOC of individual cell in battery.

Moreover, batteries are also being utilised in other modes of transportation, such as electric bicycles, scooters, and buses, further contributing to the shift towards cleaner and more sustainable transportation options. Specifically, electric bicycles, bikes, or scooters, also known as e-bikes, are a low-carbon alternative for land transport [104]. Few e-bikes are available with battery swapping technology [105], allowing for cell balancing within the e-bike and at the charging stations. In general, these types of transport mediums are available with a limited battery life. However, the battery life can be extended using various cell balancing techniques and super-capacitors [106]. Further, as e-bike technology needs many extra advanced components it incurs higher higher costs which depends on battery cost, etc. Further, name of brand also plays a key role in final price, since brands in the bike industry for years have design, reliability, and performance reputation, and may charge more.

Also, battery-powered electric ferries have emerged as a promising solution to reduce environmental impact of maritime transportation. Unlike conventional fuel-powered vessels, electric ferries rely on batteries to drive their propulsion systems, eliminating emissions of pollutants [107]. Further, these ferries are mainly advantageous for long-distance travel, which requires high autonomy. Ferries with carbon free propulsion use green hydrogen storage, and fuel cells in hybridisation with fast acting battery storage. A detailed design of these ferries accounts for perspectives related to economics, environment, and technology to find solutions to a multi-objective function. However, current designs for ferry boats assume similar power requirements as for EVs, which is a current limitation to be addressed [108]. A shift towards battery-powered propulsion of ferries represents a significant step towards achieving cleaner air quality [109].

In the railway sector, electrical traction systems stand out as the most energy-efficient option, largely due to the presence of regenerative braking systems. With regenerative brakes, traction motor functions as a generator, converting kinetic energy back into electrical energy. However, managing this regenerated energy poses the challenge that it must either be utilised immediately or stored for future use. Directly feeding energy back to power grid is an option; however, there are limitations on the receptivity of the regenerative brakes. In cases when there is no immediate demand for the regenerated energy, such as when one train accelerates while the other decelerates, excess energy is converted into heat and dissipated within the surroundings. To prevent wastage and optimise the energy usage, it is essential to store the surplus energy. This can be achieved either by storing it onboard the train or using storage devices located along the track infrastructure [93].

Overall, it is paramount to consider multiple important factors when implementing cell balancing in batteries for the transport sector. Example factors are i) BMS configuration, which supports cell balancing; ii) accurately estimating the battery parameters within a given budget; and iii) dealing with the increased complexity of cell balancing at either the individual cell or the pack level depending on type of battery chemistry. Additionally, it is important to monitor the safety of scheduled cell balancing operations. Specifically, EVs will always be prone to safety Hazards due to the overcharged cells since the batteries will have higher energy concentrated over smaller volumes. This calls for efficient battery equalisation systems since cell imbalance will result in poorer performance and many safety hazards. Also, overheating and overcharging of batteries will result in active components reacting with electrolyte, and with one other, leading to possible explosions and/or fires. It must be ensured that voltages are not extremely high as it can easily result in electrolyte decomposition. Overall, aforementioned EVs can contribute to a large extent in advancing the transport sector, and aid in reducing dependency on resources that harm human health and ecosystem.

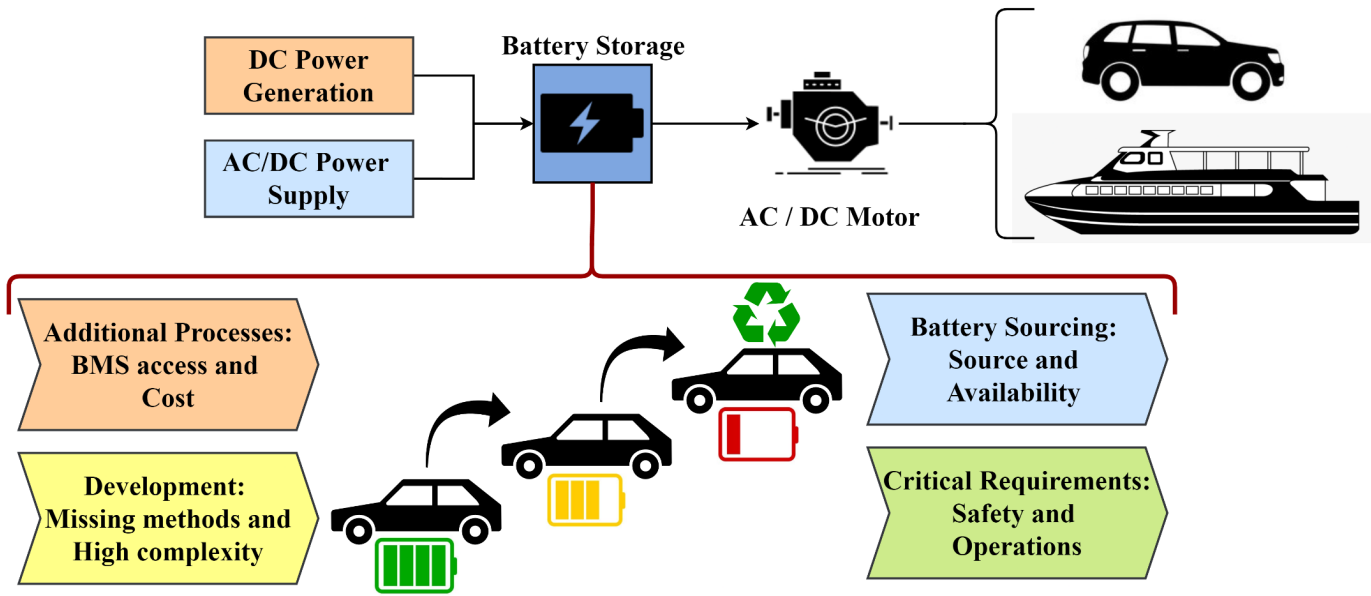

Fig. 12. Cell balancing in Transportation Sector

## V. PROSPECTS

Herein, we outline cell-balancing prospects, including a variety of research challenges with respect to standardisation, possible applications, performance, techniques and safety with their future potential road-map for research. These challenges are in detail outlined after fulfilling the various steps of data synthesis amid the survey process as mentioned in Section I. Indeed, it has been noticed that these major hindrances are stalling the growth of cell balancing as detailed next.

### A. Challenges

The cell-balancing techniques outlined in the literature are typically tailored to either small, medium, or large battery packs, yet they are often not universally compatible across all sizes. Thus, there's an opportunity to develop versatile cell-balancing techniques applicable to all types of battery packs.

This also requires safety standards and test techniques so that the batteries and the related components meet the required safety criteria. This will ensure that standardised tests and settings are followed so that the portable electronics devices and EVs, and the batteries themselves, can be upgraded. Further, the issue of thermal runaway which possess risk in commercially available devices/products can be minimised for batteries used in the EVs once they are certified following relevant safety testing standards before being produced in mass or being up for sale. In specific, standards for battery safety need to be continuously updated and optimised since the existing tests will not be able to completely guarantee battery safety for practical applications. In this regard, multiple international organisations and governments are involved in formulating and regulating battery safety following the evolving requirements and conditions of the nations. The academics and research community, together with the industry collaborators, are also involved in conducting extensive research related to battery safety. Next, we discuss the various safety standards and tests rolled out for battery safety.

In general, three principles are required to be satisfied when the safety standards are formulated. These are: (i) test operability, which refers to the fact that safety tests need to be feasible and must not include any elements which are technically incorrect or unscientific; (ii) repeatability of the test, which implies that all aspects and procedures of the test must be as consistence as possible at the same test centre, obtaining consistent results over multiple tests of any specific battery sample when employing the same experimental test-bench/equipment(s) under the identical conditions; and (iii) test reproducibility, which implies that identically correct results must be obtained, even if the test is conducted at various test centres, when employing the same experimental test-bench/equipment(s) under the identical conditions.

The most popular safety standards that have been developed thus far include the (i) Chinese standard GB/T 31,485 [110], (ii) Society of Automotive Engineers (SAE) standard 2464 [110], (iii) International Electrochemical Commission (IEC) standard IEC62133 Edition 2.0 [111] (iv) United Nations (UN) standard UN38.3 [112], (v) Japanese Industrial Standard (JIS) C8714 [113], (vi) Underwriters Laboratories (UL) standard UL2580 Edition 2.0 [114], [115], and (vii) International Standardisation Organisation (ISO) standard ISO 16750–2 [116]. These safety standards implement various technologies in regard to test requirements and parameters [110], [117]–[119]. The most popular safety test standards are summarised in Table. X.

TABLE X
INTERNATIONAL SAFETY TEST STANDARDS.

| Name | Code | Year | Characteristics | Battery Application | | | |
|---|---|---|---|---|---|---|---|
| | | | | Cell | Module | Pack | System |
| EV and HV rechargeable Energy storage system safety and abuse Testing (SAE) | SAE J2464-2009 [119] | 1999 | Requirements for safety | | | ✓ | ✓ |
| Safety requirements and test techniques for traction battery of EV (GB/T) | GB/T 31485–2015 [110] | 2015 | Requirements for environmental suitability, safety and electrical performance | ✓ | ✓ | | |
| UN manual on hazardous materials transport tests and standards, part 3, section 38.3 (UN) | UN38.3 [112] | 2015 | Requirements for safety | ✓ | ✓ | ✓ | |
| Road vehicles - environmental conditions and testing for electrical and electronic equipment - part 2: electrical loads (ISO) | ISO 16750–2 [116] | 2023 | Safety test specifications and Reliability | ✓ | ✓ | ✓ | |
| Battery safety standards for EVs (UL) | UL 2580– 2022 [114], [115] | 2022 | Requirements for environmental suitability, safety and electrical performance | ✓ | ✓ | ✓ | ✓ |
| General motors battery test standard for EVs (GM) | GMModified USABC [117] | 2016 | Requirements for abuse tolerance and electrical performance | ✓ | ✓ | | |
| VW PV8450 (VW) | VW PV8450 [118] | 2016 | Requirements for abuse test and electrical performance | ✓ | ✓ | ✓ | |

In specific, the test standards aid in minimising thermal runaway accidents' probability in practice. They have the capability of assessing the batteries' responses in real-time while simultaneously providing continuous updates and upgrades as per the evolving battery technology. Further, there are tests for assessing the overcharge and over-discharge of batteries within a cell, specifically when either of the charge or the discharge process is not in control. Then, external short-circuit tests are used for assessing short-circuiting, which is mainly due to external battery pole's electrical connections under abnormal conditions. The extreme drop test are usually initiated on the batteries to assess its responses, as batteries undergo intense rigid impacts during day to day usage or at the time of transportation. A popular test called as heating tests is usually conducted on battery to assess its thermal runaway. These thermal runaway are usually caused by heating of battery due local overheating, and which is succeeded by thermal runaway expansion. The crush tests is employed to get responses from the battery cells subjected to mechanical deformity due to an external force during a collision, which results in short-circuiting, separator rupture, damage to cathode or anode connection or both, and at the end with thermal runaway. To test the conditions of short-circuit due to the penetration of any impurities to the separator, a nail penetration test is employed. Finally, temperature shock test is used to test the responses of the batteries operating under extreme temperatures.

Once the results from the above test(s) for evaluating the batteries' safety condition are obtained, the EUCAR and the SAE-J hazard levels are widely applied [120]. In general, the Electrical Energy Storage System (EESS) devices' hazard levels are used as per responses to the conditions. In turn, this is useful for the manufacturers and integrators when they evaluate a given EESS design's response. Further, it is recommended by the SAE that test results are reported in terms of various levels of hazard severity [119], and this information must be used for battery safety and it's hazard risk migration.

The escalating electrical load demand poses a notable challenge, demanding modernisation of the current power system structure. However, economic, environmental, and technical constraints limit the extent of this modernisation effort [121]. Consequently, renewable energy sources such as solar Photovoltaic and wind are increasingly utilised to supplement the energy demand [122]. However, the erratic nature of these renewable energy sources calls for energy storage solutions, for which batteries are a popular choice. Concurrently, the expansion of electronic loads has contributed to immense power quality issues [121]. In modern electrical systems, specifically smart grids, battery cells are an integral component, as shown in Fig. 13. These battery cells serve various purposes, including storing the excess energy during times of low supply demand and supplying electricity during peak periods or when the renewable energy sources are not generating power [123]. Further, they help to maintain a balance in the supply and demand, and ensure a reliable power supply to the consumers. Additionally, battery storage systems in smart grids support grid stability by providing frequency regulation and voltage control services thereby, enhancing the grid stability, efficiency, and reliability [124], [125]. Overall, the integration of battery cells in modern electrical systems plays a decisive role in enhancing grid efficiency, reliability, and sustainability [121], and this aspect requires further investigation.

With the increasing demand for RES integration within the grid [126], driven by factors such as increasing load demand and depletion of fossil fuels, a need for storage solutions with larger capacities emerges. This often involves employing series-parallel combinations of the cells to meet the energy storage requirements effectively. While Pb-acid batteries have traditionally served as a common storage solution despite their drawbacks, such as limited energy storage capacity, limited life span, heavyweight, etc., [34], [127], there is an opportunity to explore the feasibility of adopting the Li-ion batteries for the same purpose. Additionally, research efforts towards cell balancing techniques tailored for batteries with larger capacities are needed for optimising energy storage systems.

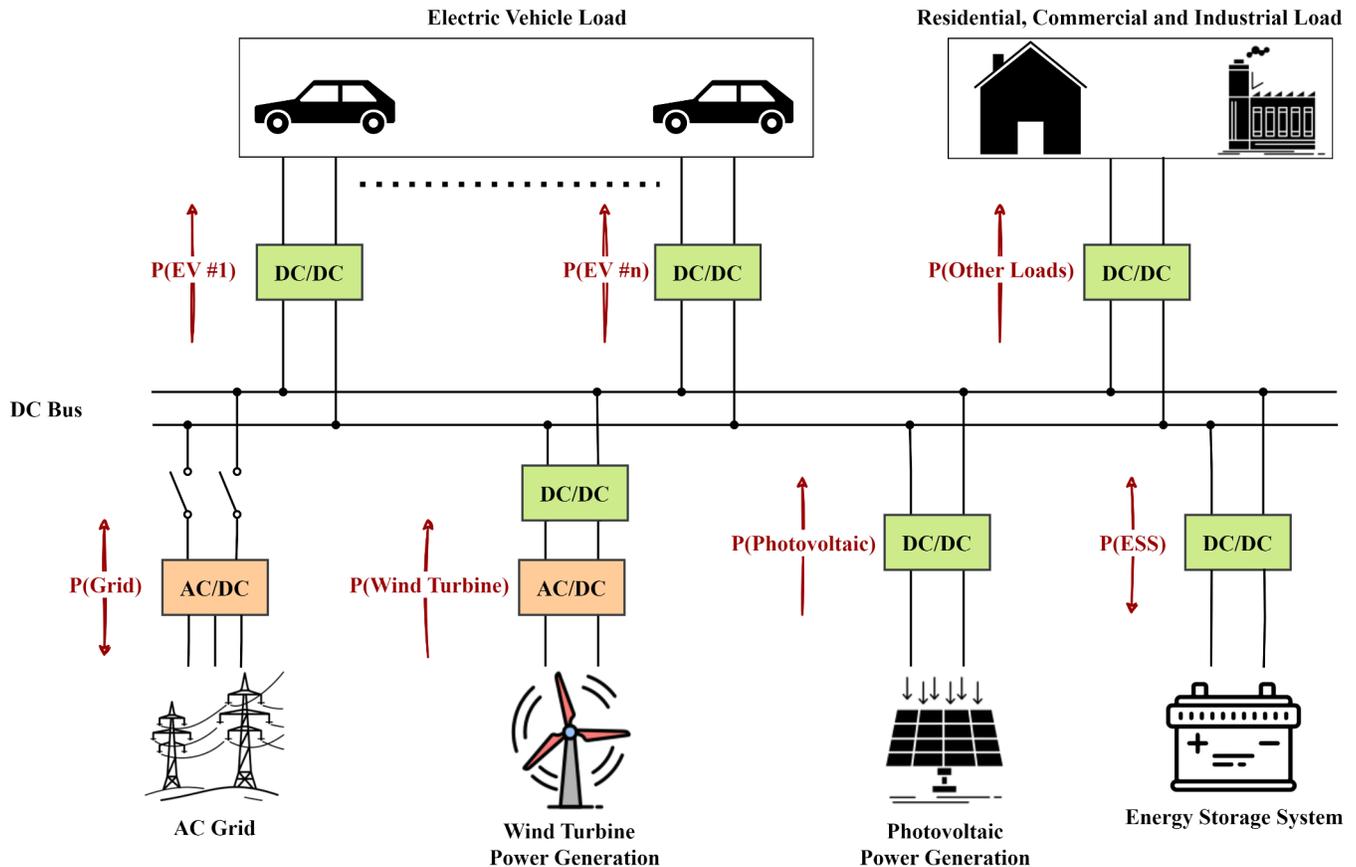

Fig. 13. Cell balancing in smart grids

Batteries are integral to numerous commercial electronic devices as shown in Fig. 14, providing the portability and convenience necessary for their operation. From smartphones to laptops, wearable fitness trackers to wireless headphones, batteries power these devices without requiring a constant external power source [128]. However, there are still multiple key issues requiring timely solutions to enhance the performance of these electronic devices. Further, it is often assumed that the BMS is well-informed about the SOE and SOC of each cell in the battery pack for effective cell balancing. However, in practical applications, the continuous monitoring and balancing cycles of each cell are often inaccurate [129] or it is impossible in most BMS either due to the cell or the BMS configuration [130]. Hence, it is challenging to continuously monitor and balance each cell, especially in commercial electronics using small-sized cells or batteries. Additionally, the heat generated by the wearables can make it difficult to accurately measure the temperature, thus SOC, SOH, and SOE parameters, of each cell. This affects the cell-balancing process, which must account for this [131].

The demand for batteries in various applications is increasing and hence there is a need for high-performance, reliable, and long-lasting battery systems for which cell balancing enabled by Artificial Intelligence (AI) can be implemented. Developing AI algorithms to anticipate cell imbalance and potential safety risks that may arise during the cell balancing process is necessary. The risks could include overheating, overcharging, over-discharging, etc. However, the key challenge will be to choose the appropriate structure for the AI techniques. Multiple hyper-parameters such as weights, batch sizes, hidden layers, biases, hidden neurons, learning rates, time steps, etc., are often used for framing complex AI based algorithms. Choosing and employing the best hyper-parameters with suitable activation functions and then training the algorithms with proper data will drastically minimise the possible computational complexities, hence solving the issues of over-fitting and under-fitting. However, to find the best suitable parameters the procedure is most time consuming with hit-and-trail tests without a proper human expertise.

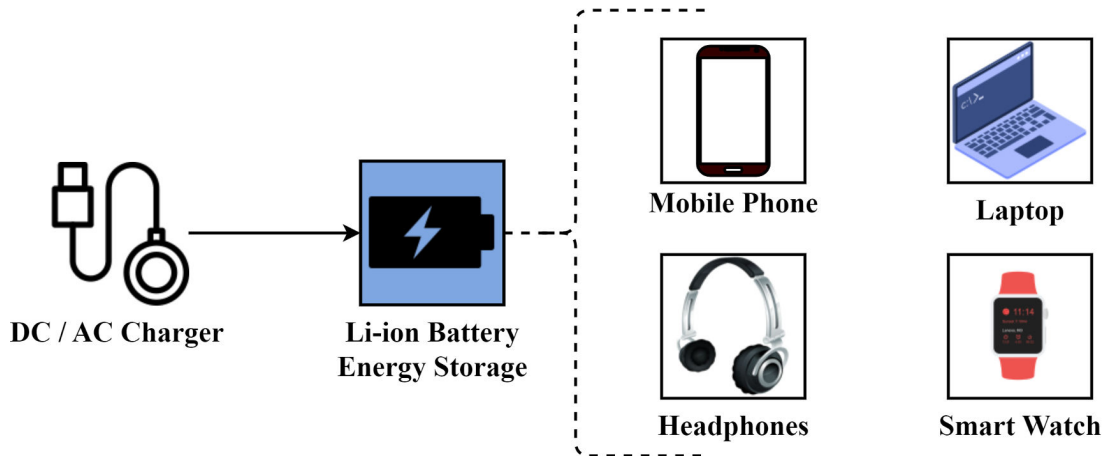

Fig. 14. Cell balancing in Commercial Electronics devices

Therefore, a generalised formulation of highly reliable and possibly precise framework is needed. This framework may aid for advanced BMS outcomes.

The BMS data quantity requirements are directly proportional to the accuracy, reliability and robustness of the AI algorithms designed for the BMS controllers. It would be an ideal path to assist the algorithms with various platforms such as big data, cloud storage and computing for enhancement in real-world situations. This will help in developing an intelligent monitoring and control unit for the battery. This unit will conduct pre-processing, and investigation of the obtained data from the battery and offer various valuable recommendations which may be opted for any future upgradations. This smartly designed unit with the Internet of Things (IoT) may also aid in the valuation, tracking and storing of SOC, SOH and other such states of batteries including fault and thermal runaway identification data for further processing in the cloud, until the end-of-life support of the battery. However, possible integration of the above-mentioned platforms with IOT may also hinder a larger scale of concerns such as respecting the privacy and security of the data, data management and scalability, interoperability, and following regulatory and legal issues. Hence, research must be directed to address these key issues.

To enhance the performance of the system, optimisation techniques, including those assisted by AI, will play a crucial role. Related research must focus on optimising cell losses, balancing speed and efficiency, system cost, etc. Note that integrating AI and conventional optimisation techniques incurs often a large amount of time while initialising parameters and then running the operational loop requires in-depth expertise. Although these issues are critical and require fundamental research, AI-assisted optimisation techniques will substantially improve the prediction efficiency, accuracy, and durability of BMS.

As highlighted in Section II, the aspects related to parallel connections require extensive research specifically directed toward the identification, analysis, and solutions of critical inconsistencies. Another concern is the case when the number of wires and connectors between the cell monitors increases. This presents two challenges viz., (i) changes in layout such as split packs increase wire length, cost, and weight, and (ii) increase in failures such as wire breaks and poor connector contact. Future research must consider these challenges during the design of BMS and cell balancing techniques. Further, the existing literature strongly supports the adoption of Li-ion batteries for EVs despite their susceptibility to thermal runaway and fires under certain conditions such as damage or overcharging. However, there is scope for exploring the use of alternatives such as Li-polymer batteries which mitigate concerns related to leakage; however, may offer slightly lower performance compared to the Li-ion variants. This requires further investigation into the feasibility and suitability of Li-polymer batteries for EV applications.

Typically, a battery in an EV consists of approximately 100 to 200 cells per pack, which are monitored and controlled by a BMS. If cells overcharge, or deep discharge, or are exposed to very high or low temperatures, or incur chemical issues, the entire battery will be unusable. Hence, the battery is considered only as good as the weakest link in the chain, which implies the weakest battery cell. This presents a key challenge in the form of cell voltages which drift apart due to manufacturing tolerances and ageing. Hence, to extend the battery lifetime, a BMS must maintain adequate balancing between the cells to recover the bad cells, simultaneously saving the good cells. To test the balancing algorithms and security features, a battery cell simulator is a key tool. This simulator must be able to emulate the cells such that they are configurable in all the characteristics. Hence, it will be required to develop such simulators to simulate all the needed states and failures.

Active battery balancing control has received extensive attention and has demonstrated promising results for reducing cell imbalance. However, research endeavours often assume specific variations among all the cell parameters, resulting in unrealistic simulation assumptions, thereby increasing the control complexity. Additionally, since there is no formal form of quantification on EV range minimisation due to cell imbalance, there are no benchmarks for evaluating the control algorithms for reducing cell imbalance, making existing balancing controllers obsolete. Hence, how cell imbalance impacts EV range is yet to be clarified. Specifically, numerical quantification minimisation on EV range, attributed majorly to cell imbalance, is a must. This aspect is addressed in [132], on implementation of an environment which is simulated with real-world vehicle dynamics data such as battery cell-level modelling, propulsion, vehicle longitudinal control and driving speed data. Specifically, to understand various key variation parameters and its impact on EVs range, equivalent circuit model is used. These equivalent circuit model is basically modelled from every battery cell. However, much more research is desired considering this aspect.

Safety plays a crucial role in various battery-related applications. There is an opportunity to explore potential faults in cell balancing circuits and conduct a comprehensive study to proactively address these issues. Incorporating protective circuits as an integral component of the cell balancing circuit is crucial. Many investigations have been conducted to realise hardware implementation with reduced circuit component count. It is worth noting that there is a trade-off between cost, balancing speed, control complexity, and overall simplicity. Further improving the design and conducting testing and validation of the hardware circuit is required to tame these trade-offs.

*B. Future Road-map*

Through this detailed comprehensive survey, we identified various key research issues related to cell balancing. Next, we propose key direction(s) in which the research can be extended.

A key identified issue concerns the creation of a standardised dataset for estimating the SoC and diagnosing faults in batteries. This aspect is crucial for ensuring the accuracy, safety, and reliability of the BMSs, especially in applications such as EVs, wherein batteries from various manufacturers with different specifications are utilised. By gathering a diverse range of data from various battery types and operating conditions, robust algorithms that accurately predict the SoC and identify potential faults in the battery cells can be formulated. Such a standardised dataset will allow for more precise modelling and analysis, resulting in improved performance and longevity of the batteries for use in real-time applications. Additionally, the readiness of a common set of data will enable the industry and academia to compare the results across different studies, which will facilitate collaboration and advancements in battery technology. Overall, creating a standardised dataset will be key in developing more efficient and effective BMSs, which will benefit both, the manufacturers and the consumers, over longer periods.

Another major issue concerns battery faults, which significantly impact the performance and lifespan of battery-operated devices. Common indications of battery faults include a reduction in runtime, difficulty in holding a charge, overheating during charging, and physical damage to the battery casing. To identify such faults accurately, professionals use specialised equipment, such as multi-meters to measure voltage levels, and examine the overall condition of the battery. Thermal imaging cameras can be utilised to detect abnormal heat signatures that may suggest internal defects or malfunctions. Regular testing and monitoring of the batteries is crucial to prevent potential safety hazards, including short circuits or explosions. By promptly identifying and addressing battery faults, professionals can ensure optimal performance and safety for both, the device and the user. Hence, it is required to design advanced techniques to timely identify and correct faults.

The survey has also revealed that implementing cell balancing for optimising battery performance is also a major hurdle. As batteries age, cells can become unbalanced, leading to reduced capacity and safety hazards. The BMSs use active or passive techniques to ensure that each cell operates within its optimal range, thereby improving energy storage capacity and lifespan. Hence, appropriate cell balancing techniques are essential for advanced applications, reducing downtime, productivity loss, and risks. The accurate implementation of cell balancing techniques can ensure safe and efficient battery operation, increasing reliability and safety.

Excessive heat generation in the batteries is also a key concern. The extemporaneous overcharging that takes place in the batteries is one of the major causes of generation and abnormal accumulation of heat. This abnormal rise in heat is succeeded by gas formation in the batteries. These conditions can be controlled by managing the voltage of the battery and its temperature. By harnessing required preventative techniques such as external overcharge protection, extemporaneous overcharging can be avoided. The mentioned preventative measure can be coupled with protection from abuses such as thermal and mechanical by elevating the outer shells of the battery. But this coupled measure comes with a drawback of high energy density hungry BMS which in response demands for intense BMS system. This drawback also results in a great extent of thermal imbalances due to immense heat emission. With more enforced protection as a solution, the heat transfer can be made more uniform and speedier in higher-density hungry batteries. Also, the installation of batteries can be done in the region of least distress at the times of collisions can be an alternative solution. The incompetence to dissipate localised heat quickly and efficiently occurs due to the failure of the separator in case of short-circuit of the battery under thermal runaway. The suggested solution can be the deployment of temperature-sensitive materials and internal fixtures of pathways for rapid diffusion of heat. Lastly, cautious selection of electrolytes, the material of active electrodes and protecting material for electrode surfaces is essential for protection from the production of heat and gas.

Regarding integrating the optimisation techniques with the AI techniques, predictions could be counterfactual due to the lack of capabilities of selection and searching of parameters. These integration concerns must be a notable point to address for further research. This stages an open research area on how these AI techniques can be implemented on a low storage memory and a low-cost processor-based real-time BMS. This open research area also comes with a requirement for an extensive examination plan for the creation of a real-time BMS-embedded system prototype. Lastly, considering environmental sustainability, a well-planned recycling or disposal system has to be developed for batteries. In an extension of the recycling or reusing of the batteries, the current research is also moving towards the creation of more environmentally friendly and sustainable new battery chemistries. However, a greater and complete approach to developing such chemistry for batteries needs to consider the holistic life cycle and the process from manufacturing to disposal of the batteries and achieve the Sustainable Development Goals (SDGs). Table. XI summarises the various issues and challenges identified through our rigorous survey and possible solutions.

TABLE XI
SUMMARY OF CHALLENGES AND PROPOSED DIRECTIONS.

| S. No. | Constraint | Existing Challenge | Proposed Directions |
|---|---|---|---|
| 1. | Lack of standardised dataset applicable across all battery types | Creating a standardised dataset for SoC estimation and fault diagnosis. | • Researchers, industry stakeholders, and regulatory bodies collaborate to establish a standardised dataset covering various battery types.<br>• The dataset should comprise detailed battery behaviour and performance data, along with information on fault characteristics.<br>• Creation of robust algorithms and diagnostic techniques. |
| 2. | Standardised dataset applicable across all battery types for battery fault identification. | Accurate identification of battery faults and creating a standardised dataset for battery fault identification. | • Collecting comprehensive data on battery behaviour and fault characteristics across diverse battery chemistries and specifications.<br>• Accurate identification of battery faults while facilitating the creation of a standardised dataset for fault identification. |
| 3. | Limited discussion on the application of cell balancing techniques in large capacity batteries for smart grid applications. | Do the current cell balancing techniques/circuits designed for EVs apply to high-capacity batteries intended for smart grid applications as well? | Evaluate the compatibility and effectiveness of existing cell balancing techniques/circuits for EVs in the context of batteries utilised for smart grid applications through rigorous testing and simulation. |
| 4. | Lack of a universal cell balancing technique applicable to all batteries irrespective of their specifications and chemistry. | The absence of a standardised cell balancing technique poses challenges for achieving optimal battery performance and longevity across different battery chemistries and specifications. | • Developing adaptive cell balancing algorithms tailored to specific battery chemistries and specifications.<br>• Collaboration among researchers, industry stakeholders, and regulatory bodies for effective cell balancing techniques. |
| 5. | Heat generation | Heat generation and accumulation inside batteries | • Improving the outer shells of the batteries to provide protection to the battery from the abuses such as thermal and mechanical.<br>• Increase uniformity of heat transfer and also speed of heat diffusion in high-density and large batteries.<br>• Installation of batteries in the region of least distress at the times of collisions. |
| 6. | Thermal Runaway | Efficient and quick dissipation of localised heat to dissipate | • Deployment of temperature-sensitive materials and internal fixtures of pathways for rapid diffusion of heat.<br>• Cautious selection of electrolytes, the material of active electrodes and protecting material for electrode surfaces. |
| 7. | Optimisation technique | Integrating optimisation technique with AI algorithms | Practicality issues related to the integration. |
| 8. | Real-time BMS | Integrating optimisation technique with AI algorithms | Practicality issues related to the integration. |
| 9. | AI techniques for real-time BMS | Large memory storage and high costs. | • Use low storage memory and a low-cost processor-based real-time BMS.<br>• Create a real-time BMS based embedded prototype system for operation and management |
| 10. | Choice of Li-ion batteries is largely influenced by their widespread endorsement. | • Mitigating vulnerability of Li-ion batteries to thermal runaway and fire risks.<br>• Determining suitability of alternatives such as Li-polymer batteries | • Conduct comparative studies between Li-ion and Li-polymer batteries considering performance, safety, and environmental impact.<br>• Investigate techniques to enhance performance of Li-polymer batteries.<br>• Explore advanced safety mechanisms. |
| 11. | Disposal and recycling of batteries | Creation of more environmentally friendly and sustainable new battery chemistries | Use techniques that consider the holistic life cycle of batteries to achieve SDGs. |

## VI. CONCLUSION

Efficient BMS is required to ensure excellent battery performance. However, batteries are prone to cell imbalance, which limits widespread applications and requires the implementation of cell balancing. In this survey, we extensively reviewed the requirements of cell balancing and corresponding techniques. Specifically, a detailed comparison of passive cell balancing techniques and the strengths and weaknesses of active cell balancing techniques was provided. Also, the applicability of cell balancing for both, series and parallel connected cells, was discussed. The need for cell balancing in commonly used battery types was examined together with their use in the transportation section. Lastly, we identified detailed prospects including various research challenges and the future directions.

Although there are numerous challenges, cell balancing has the potential to transform various sectors, including transportation. As further research is conducted and more advanced applications emerge, we anticipate more rigorous implementations of cell balancing in the future. Through this comprehensive survey, we aim to encourage additional research on critical issues related to cell balancing.

## REFERENCES

[1] A. Townsend and R. Gouws, "A comparative review of lead-acid, lithium-ion and ultra-capacitor technologies and their degradation mechanisms," *Energies*, vol. 15, no. 13, p. 4930, 2022.
[2] A. Verma and B. Singh, "A solar pv, bes, grid and dg set based hybrid charging station for uninterruptible charging at minimized charging cost," in *2018 IEEE Industry Applications Society Annual Meeting (IAS)*, 2018, pp. 1–8.
[3] S.-L. Wu, H.-C. Chen, and C.-H. Chien, "A novel active cell balancing circuit and charging strategy in lithium battery pack," *Energies*, vol. 12, no. 23, p. 4473, Nov. 2019. [Online]. Available: https://www.mdpi.com/1996-1073/12/23/4473
[4] G. Zhou, X. Zhang, K. Gao, Q. Tian, and S. Xu, "Two-mode active balancing circuit based on switched-capacitor and three-resonant-state lc units for series-connected cell strings," *IEEE Transactions on Industrial Electronics*, vol. 69, no. 5, pp. 4845–4858, 2022.

[5] M. K. Al-Smadi and J. A. Abu Qahouq, "Evaluation of current-mode controller for active battery cells balancing with peak efficiency operation," *IEEE Transactions on Power Electronics*, vol. 38, no. 2, pp. 1610–1621, 2023.

[6] M. Uzair, G. Abbas, and S. Hosain, "Characteristics of battery management systems of electric vehicles with consideration of the active and passive cell balancing process," *World Electric Vehicle Journal*, vol. 12, no. 3, p. 120, Aug. 2021. [Online]. Available: https://www.mdpi.com/2032-6653/12/3/120

[7] Q. Yu, C. Wang, J. Li, R. Xiong, and M. Pecht, "Challenges and outlook for lithium-ion battery fault diagnosis methods from the laboratory to real world applications," *eTransportation*, p. 100254, 2023.

[8] A. B. Ahmad, C. A. Ooi, and D. Ishak, "State-of-charge balancing control for optimal cell utilisation of a grid-scale three-phase battery energy storage system using hybrid modular multilevel converter topology without redundant cells," *IEEE Access*, vol. 9, pp. 53 920–53 935, 2021.

[9] e. a. Kumar, R. Ranjith, "Advances in batteries, battery modeling, battery management system, battery thermal management, soc, soh, and charge/discharge characteristics in ev applications," *IEEE Access*, vol. 11, pp. 129 335–129 352, 2023.

[10] S. Lee, M. Kim, J. W. Baek, D.-W. Kang, and J. Jung, "Enhanced switching pattern to improve cell balancing performance in active cell balancing circuit using multi-winding transformer," *IEEE Access*, vol. 8, pp. 149 544–149 554, 2020.

[11] F. S. J. Hoekstra, H. J. Bergveld, and M. C. F. Donkers, "Optimal control of active cell balancing: Extending the range and useful lifetime of a battery pack," *IEEE Transactions on Control Systems Technology*, vol. 30, no. 6, pp. 2759–2766, 2022.

[12] J. Wang, S. Dai, Y. Wei, J. Yu, Y. He, and S. Li, "Research on a novel multilayer equalization circuit for battery pack based on game theory control algorithm," in *2018 2nd IEEE Conference on Energy Internet and Energy System Integration (EI2)*, 2018, pp. 1–6.

[13] Z. B. Omariba, L. Zhang, and D. Sun, "Review of battery cell balancing methodologies for optimizing battery pack performance in electric vehicles," *IEEE Access*, vol. 7, pp. 129 335–129 352, 2019.

[14] C. Riczu and J. Bauman, "Implementation and system-level modeling of a hardware efficient cell balancing circuit for electric vehicle range extension," *IEEE Transactions on Industry Applications*, vol. 57, no. 3, pp. 2883–2895, 2021.

[15] Y. Li, J. Xu, X. Mei, and J. Wang, "A unitized multiwinding transformer-based equalization method for series-connected battery strings," *IEEE Transactions on Power Electronics*, vol. 34, no. 12, pp. 11 981–11 989, 2019.

[16] N. Samaddar, N. Senthil Kumar, and R. Jayapragash, "Passive cell balancing of li-ion batteries used for automotive applications," *Journal of Physics: Conference Series*, vol. 1716, no. 1, p. 012005, Dec. 2020. [Online]. Available: https://iopscience.iop.org/article/10.1088/1742-6596/1716/1/012005

[17] A. F. Moghaddam and A. Van Den Bossche, "An active cell equalization technique for lithium ion batteries based on inductor balancing," in *2018 9th International Conference on Mechanical and Aerospace Engineering (ICMAE)*, 2018, pp. 274–278.

[18] S. Khaleghi, M. S. Hosen, J. Van Mierlo, and M. Berecibar, "Towards machine-learning driven prognostics and health management of li-ion batteries. a comprehensive review," *Renewable and Sustainable Energy Reviews*, vol. 192, p. 114224, 2024.

[19] A. Turksoy, A. Teke, and A. Alkaya, "A comprehensive overview of the dc-dc converter-based battery charge balancing methods in electric vehicles," *Renewable and Sustainable Energy Reviews*, vol. 133, p. 110274, Nov. 2020. [Online]. Available: https://linkinghub.elsevier.com/retrieve/pii/S1364032120305633

[20] A. Carrera-Rivera, W. Ochoa, F. Larrinaga, and G. Lasa, "How-to conduct a systematic literature review: A quick guide for computer science research," *MethodsX*, vol. 9, p. 101895, 2022. [Online]. Available: https://linkinghub.elsevier.com/retrieve/pii/S2215016122002746

[21] B. Kitchenham and S. Charters, *Guidelines for performing Systematic Literature Reviews in Software Engineering*, 2nd ed., ser. EBSE Technical Report. Keele University and Durham University Joint Report, Tech. Rep. EBSE 2007-001, 07, 2007. [Online]. Available: https://legacyfileshare.elsevier.com/promis_misc/525444systematicreviewsguide.pdf

[22] M. Petticrew and H. Roberts, *Systematic Reviews in the Social Sciences: A Practical Guide*, 1st ed. Wiley, Jan. 2006. [Online]. Available: https://onlinelibrary.wiley.com/doi/book/10.1002/9780470754887

[23] M. Staples and M. Niazi, "Experiences using systematic review guidelines," *Journal of Systems and Software*, vol. 80, no. 9, p. 1425–1437, Sep. 2007. [Online]. Available: https://linkinghub.elsevier.com/retrieve/pii/S0164121206002962

[24] K. Krippendorff, *Content Analysis: An Introduction to Its Methodology*. 2455 Teller Road, Thousand Oaks California 91320: SAGE Publications, Inc., 2019. [Online]. Available: https://methods.sagepub.com/book/content-analysis-4e

[25] S. Karmakar, T. K. Bera, and A. K. Bohre, "Review on cell balancing technologies of battery management systems in electric vehicles," in *2023 IEEE IAS Global Conference on Renewable Energy and Hydrogen Technologies (GlobConHT)*. IEEE, 2023, pp. 1–5.

[26] T. Ebbs-Picken, C. M. Da Silva, and C. H. Amon, "Design optimization methodologies applied to battery thermal management systems: A review," *Journal of Energy Storage*, vol. 67, p. 107460, 2023.

[27] M. Kumar, V. K. Yadav, K. Mathuriya, and A. K. Verma, "A brief review on cell balancing for li-ion battery pack (bms)," in *2022 IEEE 10th Power India International Conference (PIICON)*. IEEE, 2022, pp. 1–6.

[28] A. H. A. AL-Jumaili, R. C. Muniyandi, M. K. Hasan, M. J. Singh, J. K. S. Paw, and M. Amir, "Advancements in intelligent cloud computing for power optimization and battery management in hybrid renewable energy systems: A comprehensive review," *Energy Reports*, vol. 10, pp. 2206–2227, 2023.

[29] T. A. Abdul-jabbar, A. Kersten, A. Mashayekh, A. A. Obed, A. J. Abid, and M. Kuder, "Efficient battery cell balancing methods for low-voltage applications: A review," in *2022 IEEE International Conference in Power Engineering Application (ICPEA)*, 2022, pp. 1–7.

[30] F. Eroğlu and A. M. Vural, "A critical review on state-of-charge balancing methods in multilevel converter based battery storage systems," in *2022 4th Global Power, Energy and Communication Conference (GPECOM)*. IEEE, 2022, pp. 14–19.

[31] M. Naguib, P. Kollmeyer, and A. Emadi, "Lithium-ion battery pack robust state of charge estimation, cell inconsistency, and balancing: Review," *IEEE Access*, vol. 9, pp. 50 570–50 582, 2021.

[32] R. Machlev, "Ev battery fault diagnostics and prognostics using deep learning: Review, challenges & opportunities," *Journal of Energy Storage*, vol. 83, p. 110614, 2024.

[33] Q. Yang, C. Xu, M. Geng, and H. Meng, "A review on models to prevent and control lithium-ion battery failures: From diagnostic and prognostic modeling to systematic risk analysis," *Journal of Energy Storage*, vol. 74, p. 109230, 2023.

[34] J. Campillo, E. Dahlquist, D. L. Danilov, N. Ghaviha, P. H. Notten, and N. Zimmerman, "Battery technologies for transportation applications," *Technologies and applications for smart charging of electric and plug-in hybrid vehicles*, pp. 151–206, 2017.

[35] Z. M. Ali, M. Calasan, S. H. A. Aleem, F. Jurado, and F. H. Gandoman, "Applications of energy storage systems in enhancing energy management and access in microgrids: A review," *Energies*, vol. 16, no. 16, p. 5930, 2023.

[36] S. M. Lukic, J. Cao, R. C. Bansal, F. Rodriguez, and A. Emadi, "Energy storage systems for automotive applications," *IEEE Transactions on industrial electronics*, vol. 55, no. 6, pp. 2258–2267, 2008.

[37] S. Ci, N. Lin, and D. Wu, "Reconfigurable battery techniques and systems: A survey," *IEEE Access*, vol. 4, pp. 1175–1189, 2016.

[38] P. Van den Bossche, J. Matheys, and J. Van Mierlo, "Battery environmental analysis," *Electric and Hybrid Vehicles: Power Sources, Models, Sustainability, Infrastructure and the Market*, p. 347, 2010.

[39] A. G. Olabi, M. A. Allam, M. A. Abdelkareem, T. Deepa, A. H. Alami, Q. Abbas, A. Alkhalidi, and E. T. Sayed, "Redox flow batteries: recent development in main components, emerging technologies, diagnostic techniques, large-scale applications, and challenges and barriers," *Batteries*, vol. 9, no. 8, p. 409, 2023.

[40] M. Skyllas-Kazacos, M. Chakrabarti, S. Hajimolana, F. Mjalli, and M. Saleem, "Progress in flow battery research and development," *Journal of the electrochemical society*, vol. 158, no. 8, p. R55, 2011.

[41] B. Scrosati and R. J. Neat, "Lithium polymer batteries," in *Applications of electroactive polymers*. Springer, 1993, pp. 182–222.

[42] M. Daowd, N. Omar, P. Bossche, and J. Van Mierlo, "Capacitor based battery balancing system," *World Electric Vehicle Journal*, vol. 5, no. 2, p. 385–393, Jun. 2012. [Online]. Available: http://www.mdpi.com/2032-6653/5/2/385

[43] L. Liu, B. Xu, Z. Yan, W. Zhou, Y. Li, R. Mai, and Z. He, "A low-cost multiwinding transformer balancing topology for retired series-connected battery string," *IEEE Transactions on Power Electronics*, vol. 36, no. 5, pp. 4931–4936, 2021.
[44] X. Luo, L. Kang, C. Lu, J. Linghu, H. Lin, and B. Hu, "An enhanced multicell-to-multicell battery equalizer based on bipolar-resonant lc converter," *Electronics*, vol. 10, no. 3, p. 293, Jan. 2021. [Online]. Available: https://www.mdpi.com/2079-9292/10/3/293
[45] P. V. Kotrannavar, Shweta, Meghana, Yashaswini, M. Kappali, and A. R. Itagi, "Enhancing the range of electric vehicles with optimum speed as a key parameter," in *2023 IEEE International Conference on Power Electronics, Smart Grid, and Renewable Energy (PESGRE)*, 2023, pp. 1–5.
[46] M. U. Ali, A. Zafar, S. H. Nengroo, S. Hussain, M. Junaid Alvi, and H.-J. Kim, "Towards a smarter battery management system for electric vehicle applications: A critical review of lithium-ion battery state of charge estimation," *Energies*, vol. 12, no. 3, p. 446, 2019.
[47] P.-H. La, T. C. Tin, and S.-J. Choi, "Dynamic resistance battery equalization for capacity optimization of parallel-connected cells," in *2019 10th International Conference on Power Electronics and ECCE Asia (ICPE 2019-ECCE Asia)*. IEEE, 2019, pp. 1–6.
[48] W. Han and L. Zhang, "Charging and discharging spaces-based current allocation in parallel-connected battery systems," in *2016 IEEE International Conference on Automation Science and Engineering (CASE)*. IEEE, 2016, pp. 1209–1214.
[49] S. W. Moore and P. J. Schneider, "A review of cell equalization methods for lithium ion and lithium polymer battery systems," in *SAE Technical Paper*, Mar. 2001, pp. 2001–01–0959. [Online]. Available: https://www.sae.org/content/2001-01-0959/
[50] R. Kallimani, S. Gulannavar, K. Pai, and P. Patil, *A Detailed Study on State of Charge Estimation Methods*. Singapore: Springer Singapore, 2022, vol. 844, p. 191–207. [Online]. Available: https://link.springer.com/10.1007/978-981-16-8862-1_14
[51] X. Cui, W. Shen, Y. Zhang, and C. Hu, "Improved active cell balancing approach based on state of charge for lithium iron phosphate batteries," in *2018 13th IEEE Conference on Industrial Electronics and Applications (ICIEA)*. IEEE, 2018, pp. 389–394.
[52] Z. Chen, W. Liao, P. Li, J. Tan, and Y. Chen, "Simple and high-performance cell balancing control strategy," *Energy Science & Engineering*, vol. 10, no. 9, pp. 3592–3601, 2022.
[53] W. Diao, M. Pecht, and T. Liu, "Management of imbalances in parallel-connected lithium-ion battery packs," *Journal of Energy Storage*, vol. 24, p. 100781, Aug. 2019. [Online]. Available: https://linkinghub.elsevier.com/retrieve/pii/S2352152X19300441
[54] R. Gogoana, M. B. Pinson, M. Z. Bazant, and S. E. Sarma, "Internal resistance matching for parallel-connected lithium-ion cells and impacts on battery pack cycle life," *Journal of Power Sources*, vol. 252, pp. 8–13, 2014.
[55] P.-H. La and S.-J. Choi, "Combined equalizer based on switch-matrix and bi-directional converter for parallel-connected battery packs in data-center or telecommunication," in *2021 International Symposium on Electrical and Electronics Engineering (ISEE)*. IEEE, 2021, pp. 244–248.
[56] M. O. Qays, Y. Buswig, M. L. Hossain, M. M. Rahman, and A. Abu-Siada, "Active cell balancing control strategy for parallelly connected lifepo4 batteries," *CSEE Journal of Power and Energy Systems*, vol. 7, no. 1, pp. 86–92, 2021.
[57] M. Dubarry, A. Devie, and B. Y. Liaw, "Cell-balancing currents in parallel strings of a battery system," *Journal of Power Sources*, vol. 321, pp. 36–46, 2016.
[58] V. B. Vulligaddala, S. Vernekar, S. Singamla, R. K. Adusumalli, V. Ele, M. Brandl, and S. M.B, "A 7-cell, stackable, li-ion monitoring and active/passive balancing ic with in-built cell balancing switches for electric and hybrid vehicles," *IEEE Transactions on Industrial Informatics*, vol. 16, no. 5, pp. 3335–3344, 2020.
[59] M. Räber, D. Hink, A. Heinzelmann, and D. O. Abdeslam, "A novel non-isolated active charge balancing architecture for lithium-ion batteries," in *2018 IEEE 27th International Symposium on Industrial Electronics (ISIE)*, 2018, pp. 471–475.
[60] R. R. Thakkar, Y. Rao, and R. R. Sawant, "Comparative performance analysis on passive and active balancing of lithium-ion battery cells," in *2021 IEEE 18th India Council International Conference (INDICON)*, 2021, pp. 1–5.
[61] S. Wang, S. Yang, W. Yang, and Y. Wang, "A new kind of balancing circuit with multiple equalization modes for serially connected battery pack," *IEEE Transactions on Industrial Electronics*, vol. 68, no. 3, pp. 2142–2150, 2021.
[62] H. S, "Overview of cell balancing methods for li-ion battery technology," *Energy Storage*, vol. 3, no. 2, p. e203, Apr. 2021. [Online]. Available: https://onlinelibrary.wiley.com/doi/10.1002/est2.203
[63] A. Khanal, A. Timilsina, B. Paudyal, and S. Ghimire, "Comparative analysis of cell balancing topologies in battery management systems," in *Proceedings of the IOE Graduate Conference, Lisbon, Portugal*, 2019, pp. 19–21.
[64] S. Kıvrak, T. Özer, Y. Oğuz, and E. B. Erken, "Battery management system implementation with the passive control method using mosfet as a load," *Measurement and Control*, vol. 53, no. 1–2, p. 205–213, Jan. 2020. [Online]. Available: http://journals.sagepub.com/doi/10.1177/0020294019883401
[65] H. Song and S. Lee, "Study on the systematic design of a passive balancing algorithm applying variable voltage deviation," *Electronics*, vol. 12, no. 12, p. 2587, Jun. 2023. [Online]. Available: https://www.mdpi.com/2079-9292/12/12/2587
[66] R. K. Vardhan, T. Selvathai, R. Reginald, P. Sivakumar, and S. Sundaresh, "Modeling of single inductor based battery balancing circuit for hybrid electric vehicles," in *IECON 2017 - 43rd Annual Conference of the IEEE Industrial Electronics Society*, 2017, pp. 2293–2298.
[67] W. Chen, J. Liang, Z. Yang, and G. Li, "A review of lithium-ion battery for electric vehicle applications and beyond," *Energy Procedia*, vol. 158, p. 4363–4368, Feb. 2019. [Online]. Available: https://linkinghub.elsevier.com/retrieve/pii/S1876610219308215
[68] S. Singirikonda and Y. Obulesu, "Active cell voltage balancing of electric vehicle batteries by using an optimized switched capacitor strategy," *Journal of Energy Storage*, vol. 38, p. 102521, Jun. 2021. [Online]. Available: https://linkinghub.elsevier.com/retrieve/pii/S2352152X21002693
[69] F. Liu, R. Zou, and Y. Liu, "An any-cell-to-any-cell battery equalizer based on half-bridge lc converter," *IEEE Transactions on Power Electronics*, vol. 38, no. 4, pp. 4218–4223, 2023.
[70] K. Liu, Z. Yang, X. Tang, and W. Cao, "Automotive battery equalizers based on joint switched-capacitor and buck-boost converters," *IEEE Transactions on Vehicular Technology*, vol. 69, no. 11, pp. 12 716–12 724, 2020.
[71] D. Thiruvonasundari and K. Deepa, "Optimized passive cell balancing for fast charging in electric vehicle," *IETE Journal of Research*, vol. 69, no. 4, p. 2089–2097, May 2023. [Online]. Available: https://www.tandfonline.com/doi/full/10.1080/03772063.2021.1886604
[72] S. Apipatsakul, M. Masomtob, and N. Fuengwarodsakul, "On a design of adjustable passive balancing circuit using pwm technique for li-ion battery," in *2019 Research, Invention, and Innovation Congress (RI2C)*, 2019, pp. 1–5.
[73] T. Duraisamy and D. Kaliyaperumal, "Machine learning-based optimal cell balancing mechanism for electric vehicle battery management system," *IEEE Access*, vol. 9, pp. 132 846–132 861, 2021.
[74] R. Di Fonso, X. Sui, A. B. Acharya, R. Teodorescu, and C. Cecati, "Multidimensional machine learning balancing in smart battery packs," in *IECON 2021 – 47th Annual Conference of the IEEE Industrial Electronics Society*, 2021, pp. 1–6.
[75] A. Ziegler, D. Oeser, T. Hein, D. Montesinos-Miracle, and A. Ackva, "Reducing cell to cell variation of lithium-ion battery packs during operation," *IEEE Access*, vol. 9, pp. 24 994–25 001, 2021.
[76] Amin, K. Ismail, A. Nugroho, and S. Kaleg, "Passive balancing battery management system using mosfet internal resistance as balancing resistor," in *2017 International Conference on Sustainable Energy Engineering and Application (ICSEEA)*, 2017, pp. 151–155.
[77] W. Han, C. Zou, L. Zhang, Q. Ouyang, and T. Wik, "Near-fastest battery balancing by cell/module reconfiguration," *IEEE Transactions on Smart Grid*, vol. 10, no. 6, pp. 6954–6964, 2019.
[78] J. P. D. Miranda, L. A. M. Barros, and J. G. Pinto, "A review on power electronic converters for modular bms with active balancing," *Energies*, vol. 16, no. 7, 2023. [Online]. Available: https://www.mdpi.com/1996-1073/16/7/3255
[79] T. Conway, "An isolated active balancing and monitoring system for lithium ion battery stacks utilizing a single transformer per cell," *IEEE Transactions on Power Electronics*, vol. 36, no. 4, pp. 3727–3734, 2021.
[80] Y. Cao, K. Li, and M. Lu, "Balancing method based on flyback converter for series-connected cells," *IEEE Access*, vol. 9, pp. 52 393–52 403, 2021.

[81] I. Bistritz and N. Bambos, "Consensus-based stochastic control for model-free cell balancing," *IEEE Transactions on Control of Network Systems*, vol. 8, no. 3, pp. 1139–1150, 2021.

[82] S. K. Dam and V. John, "Low-frequency selection switch based cell-to-cell battery voltage equalizer with reduced switch count," *IEEE Transactions on Industry Applications*, vol. 57, no. 4, pp. 3842–3851, 2021.

[83] B. Yildirim, M. Elgendy, A. Smith, and V. Pickert, "Evaluation and comparison of battery cell balancing methods," in *2019 IEEE PES Innovative Smart Grid Technologies Europe (ISGT-Europe)*, 2019, pp. 1–5.

[84] X. Qi, Y. Wang, M. Fang, Y. Wang, and Z. Chen, "Principle and topology derivation of integrated cascade bidirectional converters for centralized charge equalization systems," *IEEE Transactions on Power Electronics*, vol. 37, no. 2, pp. 1852–1869, 2022.

[85] Q. Ouyang, Y. Zhang, N. Ghaeminezhad, J. Chen, Z. Wang, X. Hu, and J. Li, "Module-based active equalization for battery packs: A two-layer model predictive control strategy," *IEEE Transactions on Transportation Electrification*, vol. 8, no. 1, pp. 149–159, 2022.

[86] M. Raeber, A. Heinzelmann, and D. O. Abdeslam, "Analysis of an active charge balancing method based on a single nonisolated dc/dc converter," *IEEE Transactions on Industrial Electronics*, vol. 68, no. 3, pp. 2257–2265, 2021.

[87] M. Uno and K. Yoshino, "Modular equalization system using dual phase-shift-controlled capacitively isolated dual active bridge converters to equalize cells and modules in series-connected lithium-ion batteries," *IEEE Transactions on Power Electronics*, vol. 36, no. 3, pp. 2983–2995, 2021.

[88] P.-H. La and S.-J. Choi, "Reactive balancing circuit for paralleled battery modules employing dynamic capacitance modulation," in *2020 IEEE Energy Conversion Congress and Exposition (ECCE)*. IEEE, 2020, pp. 575–581.

[89] W. Wang and M. Preindl, "Dual cell links for battery-balancing auxiliary power modules: A cost-effective increase of accessible pack capacity," *IEEE Transactions on Industry Applications*, vol. 56, no. 2, pp. 1752–1765, 2020.

[90] Y. Yu, R. Saasaa, A. A. Khan, and W. Eberle, "A series resonant energy storage cell voltage balancing circuit," *IEEE Journal of Emerging and Selected Topics in Power Electronics*, vol. 8, no. 3, pp. 3151–3161, 2020.

[91] L. Liu, W. Sun, P. Han, R. Mai, Z. He, and W. Li, "Design of zero-current parallel-switched-capacitor voltage equalizer for battery strings," in *2019 IEEE Applied Power Electronics Conference and Exposition (APEC)*, 2019, pp. 3180–3183.

[92] X. Ding, D. Zhang, J. Cheng, B. Wang, Y. Chai, Z. Zhao, R. Xiong, and P. C. K. Luk, "A novel active equalization topology for series-connected lithium-ion battery packs," *IEEE Transactions on Industry Applications*, vol. 56, no. 6, pp. 6892–6903, 2020.

[93] N. Ghaviha, J. Campillo, M. Bohlin, and E. Dahlquist, "Review of application of energy storage devices in railway transportation," *Energy Procedia*, vol. 105, pp. 4561–4568, 2017.

[94] M. Ehsani, K. V. Singh, H. O. Bansal, and R. T. Mehrjardi, "State of the art and trends in electric and hybrid electric vehicles," *Proceedings of the IEEE*, vol. 109, no. 6, pp. 967–984, 2021.

[95] S. Vishnu, K. Pai, P. S. Praveena Krishna, N. S. Jayalakshmi, S. D. Suraj, and V. Prathimala, "Correlative analysis of dynamic behaviour of lithium-ion cell using matlab and typhoon hil," in *2021 IEEE International Conference on Distributed Computing, VLSI, Electrical Circuits and Robotics (DISCOVER)*, 2021, pp. 225–230.

[96] A. Verma and B. Singh, "Integration of solar pv-wecs and dg set for ev charging station," in *2020 IEEE International Conference on Power Electronics, Smart Grid and Renewable Energy (PESGRE2020)*. IEEE, 2020, pp. 1–6.

[97] T. Duraisamy and K. Deepa, "Evaluation and comparative study of cell balancing methods for lithium-ion batteries used in electric vehicles," *International Journal of Renewable Energy Development*, vol. 10, no. 3, p. 471–479, Aug. 2021. [Online]. Available: https://ijred.cbiore.id/index.php/ijred/article/view/34484

[98] S. Baccari, M. Tipaldi, and V. Mariani, "Deep reinforcement learning for cell balancing in electric vehicles with dynamic reconfigurable batteries," *IEEE Transactions on Intelligent Vehicles*, pp. 1–12, 2024.

[99] S. V. P. Singh and P. Agnihotri, "Ann based modelling of optimal passive cell balancing," in *2022 22nd National Power Systems Conference (NPSC)*, 2022, pp. 326–331.

[100] J. Gallardo-Lozano, E. Romero-Cadaval, M. I. Milanes-Montero, and M. A. Guerrero-Martinez, "A novel active battery equalization control with on-line unhealthy cell detection and cell change decision," *Journal of Power Sources*, vol. 299, p. 356–370, Dec. 2015. [Online]. Available: https://linkinghub.elsevier.com/retrieve/pii/S0378775315302676

[101] Y. Barsukov *et al.*, "Battery cell balancing: What to balance and how," *Texas Instruments*, pp. 2–1, 2009.

[102] Y. Xing, E. W. M. Ma, K. L. Tsui, and M. Pecht, "Battery management systems in electric and hybrid vehicles," *Energies*, vol. 4, no. 11, p. 1840–1857, Oct. 2011. [Online]. Available: http://www.mdpi.com/1996-1073/4/11/1840

[103] Y. Weng and C. Ababei, "Ai-assisted reconfiguration of battery packs for cell balancing to extend driving runtime," *Journal of Energy Storage*, vol. 84, p. 110853, Apr. 2024. [Online]. Available: https://linkinghub.elsevier.com/retrieve/pii/S2352152X24004377

[104] L. Stilo, D. Segura-Velandia, H. Lugo, P. P. Conway, and A. A. West, "Electric bicycles, next generation low carbon transport systems: A survey," *Transportation Research Interdisciplinary Perspectives*, vol. 10, p. 100347, Jun. 2021. [Online]. Available: https://linkinghub.elsevier.com/retrieve/pii/S2590198221000543

[105] K. Koirala, M. Tamang, and Shabbiruddin, "Planning and establishment of battery swapping station - a support for faster electric vehicle adoption," *Journal of Energy Storage*, vol. 51, p. 104351, Jul. 2022. [Online]. Available: https://linkinghub.elsevier.com/retrieve/pii/S2352152X22003759

[106] N. Hatwar, A. Bisen, H. Dodke, A. Junghare, and M. Khanapurkar, "Design approach for electric bikes using battery and super capacitor for performance improvement," in *16th International IEEE Conference on Intelligent Transportation Systems (ITSC 2013)*, 2013, pp. 1959–1964.

[107] H. Bach, A. Bergek, O. Bjørgum, T. Hansen, A. Kenzhegaliyeva, and M. Steen, "Implementing maritime battery-electric and hydrogen solutions: A technological innovation systems analysis," *Transportation Research Part D: Transport and Environment*, vol. 87, p. 102492, Oct. 2020. [Online]. Available: https://linkinghub.elsevier.com/retrieve/pii/S1361920920306799

[108] E. Zafeiratou and C. Spataru, "Modelling electric vehicles uptake on the greek islands," *Renewable and Sustainable Energy Transition*, vol. 2, p. 100029, Aug. 2022. [Online]. Available: https://linkinghub.elsevier.com/retrieve/pii/S2667095X22000137

[109] M. Cha, S. G. Jayasinghe, H. Enshaei, R. Islam, A. Abeysiriwardhane, and S. Alahakoon, "Power management optimisation of a battery/fuel cell hybrid electric ferry," in *2021 31st Australasian Universities Power Engineering Conference (AUPEC)*. IEEE, 2021, pp. 1–6.

[110] Z. Zhang, L. Zhang, L. Hu, and C. Huang, "Active cell balancing of lithium-ion battery pack based on average state of charge," *International Journal of Energy Research*, vol. 44, no. 4, p. 2535–2548, Mar. 2020. [Online]. Available: https://onlinelibrary.wiley.com/doi/10.1002/er.4876

[111] I. E. Commission, "Secondary cells and batteries containing alkaline or other non-acid electrolytes - safety requirements for portable sealed secondary cells, and for batteries made from them, for use in portable applications - part 2: Lithium systems." [Online]. Available: https://webstore.iec.ch/publication/70017

[112] U. Nations, "Recommendations on the transport of dangerous goods: Manual of tests and criteria," 2015. [Online]. Available: https://unece.org/fileadmin/DAM/trans/danger/ST_SG_AC.10_11_Rev6_E_WEB_-With_corrections_from_Corr.1.pdf

[113] J. I. S. [jis], "Jis c 8714:2007 - safety tests for portable lithium ion secondary cells and batteries for use in portable electronic applications." [Online]. Available: https://webstore.ansi.org/standards/jis/jis87142007

[114] R. Chen, Q. Li, X. Yu, L. Chen, and H. Li, "Approaching practically accessible solid-state batteries: Stability issues related to solid electrolytes and interfaces," *Chemical Reviews*, vol. 120, no. 14, p. 6820–6877, Jul. 2020. [Online]. Available: https://pubs.acs.org/doi/10.1021/acs.chemrev.9b00268

[115] U. L. I. (UL), "Ul standard for safety batteries for use in electric vehicles," 2022.

[116] ISO, "Iso 16750-2:2023 road vehicles environmental conditions and testing for electrical and electronic equipment part 2: Electrical loads," 2023. [Online]. Available: https://www.iso.org/standard/76119.html

[117] U. S. A. B. C. L. (USABC), "United states advanced battery consortium llc (usabc)." [Online]. Available: https://uscar.org/usabc/

[118] A. T. Services, "Volkswagen sae standards and testing." [Online]. Available: https://atslab.com/pdf/automotive-specs/volkswagen-specs.pdf

[119] H. E. Committee, *Electric Vehicle Battery Abuse Testing*. SAE International, 1999. [Online]. Available: https://www.sae.org/content/j2464_199903

[120] D. H. Doughty and C. C. Crafts, "Freedomcar :electrical energy storage system abuse test manual for electric and hybrid electric vehicle applications." *OSTI.GOV*, 8 2006. [Online]. Available: https://www.osti.gov/biblio/889934

[121] P. F. Ribeiro, B. K. Johnson, M. L. Crow, A. Arsoy, and Y. Liu, "Energy storage systems for advanced power applications," *Proceedings of the IEEE*, vol. 89, no. 12, pp. 1744–1756, 2001.

[122] S. Singh, B. Singh, B. Panigrahi *et al.*, "Robust control technique of grid synchronization of spv based bes system for power quality improvement," in *2020 3rd International Conference on Energy, Power and Environment: Towards Clean Energy Technologies*. IEEE, 2021, pp. 1–6.

[123] K. M. Tan, T. S. Babu, V. K. Ramachandaramurthy, P. Kasinathan, S. G. Solanki, and S. K. Raveendran, "Empowering smart grid: A comprehensive review of energy storage technology and application with renewable energy integration," *Journal of Energy Storage*, vol. 39, p. 102591, 2021.

[124] A. Reindl, H. Meier, and M. Niemetz, "Scalable, decentralized battery management system based on self-organizing nodes," in *International Conference on Architecture of Computing Systems*. Springer, 2020, pp. 171–184.

[125] A. R. Itagi, M. Kappali, S. Karajgi, and P. Kallimani, "Prediction of solar insolation in a pv based dc micro grid to meet the ceaseless energy demand of critical loads," in *2022 International Conference for Advancement in Technology (ICONAT)*. IEEE, 2022, pp. 1–3.

[126] M. Abbas, I. Cho, and J. Kim, "Scalable constrained attributes/issues about risk, reliability and optimization in large scale battery pack," *Journal of Energy Storage*, vol. 39, p. 102632, 2021.

[127] C. F. Brasil and C. L. Melo, "A comparative study of lead-acid batteries and lithium iron phosphate batteries used in microgrid systems," in *2017 Brazilian Power Electronics Conference (COBEP)*. IEEE, 2017, pp. 1–7.

[128] M. Yusof, S. Toha, N. Kamisan, N. Hashim, and M. Abdullah, "Battery cell balancing optimisation for battery management system," in *IOP Conference Series: Materials Science and Engineering*, vol. 184, no. 1. IOP Publishing, 2017, p. 012021.

[129] A. K. M. A. Habib, M. K. Hasan, G. F. Issa, D. Singh, S. Islam, and T. M. Ghazal, "Lithium-ion battery management system for electric vehicles: Constraints, challenges, and recommendations," *Batteries*, vol. 9, no. 3, p. 152, Feb. 2023. [Online]. Available: https://www.mdpi.com/2313-0105/9/3/152

[130] M. K. S. Verma, S. Basu, R. S. Patil, K. S. Hariharan, S. P. Adiga, S. M. Kolake, D. Oh, T. Song, and Y. Sung, "On-board state estimation in electrical vehicles: Achieving accuracy and computational efficiency through an electrochemical model," *IEEE Transactions on Vehicular Technology*, vol. 69, no. 3, pp. 2563–2575, 2020.

[131] A. Samanta, S. S. Williamson, "A comprehensive review of lithium-ion cell temperature estimation techniques applicable to health-conscious fast charging and smart battery management systems," *Energies*, vol. 14, no. 18, p. 5960, Sep. 2021. [Online]. Available: https://www.mdpi.com/1996-1073/14/18/5960

[132] J. Chen, Z. Zhou, Z. Zhou, X. Wang, and B. Liaw, "Impact of battery cell imbalance on electric vehicle range," *Green Energy and Intelligent Transportation*, vol. 1, no. 3, p. 100025, Dec. 2022. [Online]. Available: https://linkinghub.elsevier.com/retrieve/pii/S2773153722000251

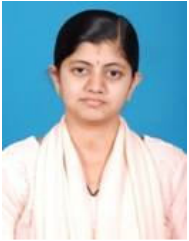

**Anupama R Itagi** received her M.Tech degree in VLSI and Embedded Systems from B V Bhoomraddi College of Engineering and Technology, affiliated with Visvesvaraya Technological University, in 2012. She completed her B.E. degree in Electrical and Electronics Engineering from SDM College of Engineering and Technology, Dharwad, in 2007. She has served as a Lecturer at BVBCET, Hubballi from 2008-2012. Currently, she is an Assistant Professor at EEE department, KLE Technological University, BVB campus, Hubballi. Her areas of interest include Renewable energy, Machine Learning, Embedded Systems, Electrical Vehicles, Battery Management Systems, and Engineering Education.

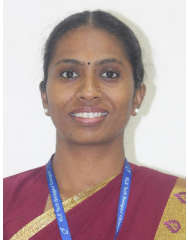

**Rakhee Kallimani** (Student Branch Counsellor and Member, IEEE) holds a Ph.D. from Visvesvaraya Technological University, which she received in 2023. She also holds an M.Tech degree in Embedded Systems from Calicut University, which she received in 2007, and a B.E. degree in Electrical and Electronics Engineering from B V Bhoomraddi College of Engineering and Technology, affiliated with Visvesvaraya Technological University, which she received in 2003. She has served as a Lecturer at BVBCET, Hubballi from 2003-2009 and as an Assistant Professor at KLE MSSCET, Belagavi from 2009 to 2023. Her areas of expertise include Embedded Systems, Wireless Sensor Networks, Electrical Vehicles, Battery Management Systems, and Engineering Education. Currently, she is an Associate Professor and also serves as the Head of the EEE department at KLE Technological University's Dr M S Sheshgiri College of Engineering and Technology in Belagavi.

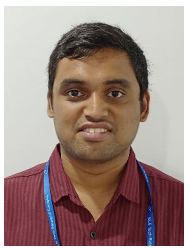

**Krishna Pai** received a Bachelor's degree in Electrical and Electronics Engineering from KLE Dr. M.S. Sheshgiri College of Engineering and Technology, Belagavi, India in 2021. He comes from an industrial background and works as an independent researcher in Bengaluru, India, to pursue his research interests. His research interests include Machine Learning, the Internet of Things, Electric Vehicles, Battery Management systems, Embedded Systems and Communication Theory. Currently, he is involved in multiple research projects, and his work encompasses over 16+ published articles in top-tier journals and conferences.

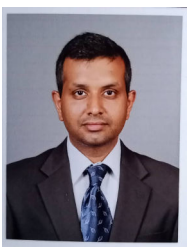

**Sridhar Iyer** (Senior Member, IEEE) received the M.S. degree in Electrical Engineering from New Mexico State University, U.S.A in 2008, and the Ph.D. degree from Delhi University, India in 2017. He received the young scientist award from the DST/SERB, Govt. of India in 2013, and Young Researcher Award from Institute of Scholars in 2021. He is the Recipient of the 'Protsahan Award' from IEEE ComSoc, Bangalore Chapter as a recognition to his contributions towards paper published/tutorial offered in recognised conferences/journals, during Jan 2020-Sep 2021, and during Oct 2021-Oct 2022). He has completed two funded research projects as the Principal Investigator and is currently involved in on-going funded research projects as the Principal Investigator. His current research focus includes semantic communications and spectrum enhancement techniques for Intelligent wireless networks. Currently, he serves as the Professor at KLE Technological University, Dr MSSCET, Belagavi, Karnataka, India.

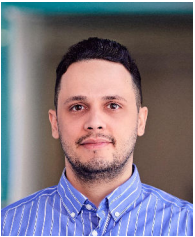
**Onel L. A. López** (Member, IEEE) received the B.Sc. (1st class honors, 2013), M.Sc. (2017) and D.Sc. (with distinction, 2020) degree in Electrical Engineering from the Central University of Las Villas (Cuba), the Federal University of Paraná (Brazil) and the University of Oulu (Finland), respectively. From 2013-2015 he served as a specialist in telematics at the Cuban telecommunications company (ETECSA). He is a collaborator to the 2016 Research Award given by the Cuban Academy of Sciences, a co-recipient of the 2019 and 2023 IEEE European Conference on Networks and Communications (EuCNC) Best Student Paper Award, the recipient of both the 2020 best doctoral thesis award granted by Academic Engineers and Architects in Finland TEK and Tekniska Föreningen i Finland TFiF in 2021 and the 2022 Young Researcher Award in the field of technology in Finland. He is co-author of the books entitled "Wireless RF Energy Transfer in the massive IoT era: towards sustainable zero-energy networks", Wiley, 2021, and "Ultra-Reliable Low-Latency Communications: Foundations, Enablers, System Design, and Evolution Towards 6G", Now Publishers, 2023. He currently holds an Assistant Professorship (tenure track) in sustainable wireless communications engineering in the Centre for Wireless Communications (CWC), Oulu, Finland. His research interests include sustainable IoT, energy harvesting, wireless RF energy transfer, wireless connectivity, machine-type communications, and cellular-enabled positioning systems.